\documentclass[aps,prb,reprint,superscriptaddress]{revtex4-2}
\usepackage{amssymb,amsmath,bm,braket,graphicx,enumerate,mathtools}
\usepackage[justification=Justified, format=plain, font=small]{caption}
\usepackage[justification=Justified, format=plain, font=small]{subcaption}
\usepackage[bookmarks=true,colorlinks,citecolor=blue,urlcolor=blue]{hyperref}
\usepackage{pgfplots}
\usepackage{tikz,float}

\usetikzlibrary{calc}
\usepackage[utf8]{inputenc}
\usepackage[T2A]{fontenc}

\usepackage{comment}
\usepackage[normalem]{ulem}

\inputencoding{utf8}

\usetikzlibrary{arrows.meta, positioning}

\usepgfplotslibrary{colorbrewer}
\usetikzlibrary{matrix} % <- added for self made legend
\pgfplotsset{compat=1.18}

\begin{document}

% \title{2D\,+1 Non-Ergodic Dynamics in Classical and Quantum Parity-Check Circuits}

%\title{Circuits built from Pairwise Difference Conserving Gates: 
%systems with an extensive amount of local loop charges}

\title{Loop Charges and Fragmentation in Pairwise Difference Conserving Circuits}

\newcommand{\ULFMF}{Department of Physics, Faculty of Mathematics and Physics, 
University of Ljubljana, Jadranska 21, SI-1000 Ljubljana, Slovenia}

\author{Pavel Orlov}
\affiliation{\ULFMF}
\affiliation{Nanocenter CENN, Jamova 39, SI-1000 Ljubljana, Slovenia}
\author{Cheryne Jonay}\affiliation{\ULFMF}
\author{Tomaž Prosen}\affiliation{\ULFMF}
\affiliation{Institute of Mathematics, Physics and Mechanics, Jadranska 19, SI-1000 Ljubljana, Slovenia}
%\date{May 2025}

\begin{abstract}
In this work, we introduce a broad class of circuits, or quantum cellular automata, which we call ``pairwise-difference-conserving circuits'' (PDC). These models are characterized by local gates that preserve the pairwise difference of local operators (e.g. particle number). Such circuits can be defined on arbitrary graphs in arbitrary dimensions for both quantum and classical %local 
degrees of freedom. 
A key consequence of the PDC construction is the emergence of an extensive set of loop charges associated with %arbitrary 
closed walks of even length on the graph. These charges exhibit a one-dimensional character reminiscent of 1-form symmetries and lead to strong Hilbert-space fragmentation. As a case study, we analyze a quasi one-dimensional ladder geometry, where we characterize all dynamically disconnected sectors by the loop-charge symmetries, providing a complete decomposition of the Hilbert space. For the ladder geometry, we observe clear signatures of nonergodic dynamics even within the largest symmetry sector.

% Within the largest sector, dynamics appear ergodic, yet a subset of special initial states displays pronounced late-time revivals.% via a mechanism distinct from quantum scarring.
\end{abstract}

\maketitle
%\tableofcontents

\section{Introduction}

Local symmetries play a central role in understanding the dynamics and statistical properties of many-body systems, both classical and quantum. In particular, symmetries can strongly restrict the motion of the system, giving rise to a variety of nonergodic behaviors.

A paradigmatic example are integrable systems in one spatial dimension~\cite{Franchini_2017, korepin1993quantuminversescatteringmethod, Caux_2011}, whose dynamics are governed by an extensive number of local or quasilocal conserved charges~\cite{Ilievski_2016}. These constraints prevent thermalization in the conventional sense and lead to relaxation towards generalized Gibbs ensembles~\cite{Caux_2013, Ilievski_2015, Vidmar_2016, Essler_2016}. Beyond stationary properties, the presence of an extensive number of local conservation laws gives rise to a rich hydrodynamic description of non-equilibrium transport known as generalized hydrodynamics (GHD)~\cite{Castro_Alvaredo_2016, Bertini_2016, Essler_2023, Doyon_2020}.

A different route to ergodicity breaking arises in putative many-body localized (MBL) phases, where large enough quenched disorder prevents thermalization even in the presence of interactions~\cite{Abanin_2019, Sierant_2025}. In such systems, the dynamics are governed by an extensive set of emergent local integrals of motion (LIOMs) which remain exponentially localized in space and ensure that local observables retain memory of the initial state, even at asymptotically long times~\cite{Serbyn_2013, Imbrie_2017}.

Even in clean systems without disorder, nonergodic dynamics can emerge from Hilbert space fragmentation, as prominently observed in kinetically constrained systems~\cite{Khemani_2020, DCmodels, zadnik2021folded, Moudgalya_2022}. In such fragmented systems, local conservation laws or dynamical selection rules prevent transitions between different sectors of the Hilbert space, partitioning it into exponentially many disconnected subspaces (strong fragmentation has no dominant sector; weak fragmentation has a single sector capturing almost all states). Such constraints cause slow dynamics and ergodicity breaking in otherwise clean, interacting systems~\cite{Khemani_2020, zadnik2023slow, marić2025slow}.

While most studies of nonergodic dynamics have focused on one-dimensional systems, the situation in higher dimensions is far less understood. Quantum integrability has no straightforward generalization beyond one dimension, and the
many-body localized phase has been shown to be thermodynamically unstable for $d>1$~\cite{De_Roeck_2017,Sierant_2025}. Recently, however, signatures of Hilbert space fragmentation have been identified in higher-dimensional models, suggesting that constrained dynamics and nonergodicity persists even beyond one dimension~\cite{Khudorozhkov_2022,ciavarella2025,adler2024,kwan2023,stahl2025}.

In this work, our goal is to construct and analyze a broad class of models that exhibit an extensive number of local conservation laws and can be consistently defined in arbitrary spatial dimensions. To this end, we introduce pairwise-difference-conserving (PDC) circuits, which are local quantum circuits built out of gates that,
%a subclass of local quantum cellular automata
%built from gates that preserve, 
for every pair of sites they act on, preserve the difference of a chosen single-site observable (e.g., particle number or spin projection). These circuits can be defined on graphs of arbitrary dimension, and for both classical and quantum degrees of freedom.

The defining property of PDC gates naturally gives rise to a rich algebraic structure of local loop symmetries. In particular, we identify a set of conserved loop charges, associated with arbitrary closed walks of even length on the underlying graph. These charges display a one-dimensional character representing a lattice version of 1-form symmetries~\cite{Gaiotto_2015,McGreevy_2023}, and strongly fragment the Hilbert space. %their presence strongly constrains the accessible Hilbert space, leading to Hilbert-space fragmentation into exponentially many dynamically disconnected subsectors.

To characterize the resulting fragmentation, we focus on a quasi one-dimensional ladder geometry, which provides a minimal setting featuring nontrivial loops. Within this geometry, we demonstrate that all dynamically disconnected sectors of the Hilbert space can be exactly classified in terms of the underlying PDC symmetries, yielding a complete decomposition of the total Hilbert space.

Finally, for a ladder geometry we examine the dynamics within the largest (maximal) symmetry sector, where all loop symmetries have been resolved. Even in this fully symmetric sector, we observe clear signatures of nonergodic behavior, including a violation of the diagonal Eigenstate Thermalization Hypothesis (ETH) and persistent periodic revivals for a subset of product states.
%These features, however, may be specific only to this one-dimensional case.
But, while the Hilbert space fragmentation arising from PDC symmetries is universal across all graphs and dimensions, the dynamical properties within the maximal sector may depend on the specific model parameters and geometry.

\section{Definition of the circuit}

\subsection{Pairwise Difference Conserving (PDC) Gates } 

In this work, we consider quantum circuits constructed from a particular type of local gates. Let $U^{(n)}$ denote the $n$-body gate that acts on states of $n$-qubits $ \ket{ s_{1} \ldots s_n }$ with $s_{i} \in\{ 0,1\}$ \footnote{Note, however, that anything discussed below can be generalized straightforwardly to qudits of arbitrary local dimension}. We call the gate $U^{(n)}$ Pairwise Difference Conserving (PDC) if there exists a non-degenerate 1-qubit operator $A$ such that for any pair of qubits $\forall i,j \in \{1, \ldots n \}$ the gate commutes with the corresponding difference operator:
\begin{equation}\label{PDC-condition}
    [ U^{(n)} , A_{i} - A_{j}] = 0.
\end{equation}
As we discuss in the next subsection, this property is crucial for constructing local integrals of motion in circuits made from such gates. However, before proceeding in this direction, let us parametrize the most general form of gates satisfying the PDC condition~(\ref{PDC-condition}).%in the case of qubits.

%It is clear that for qubits, 
Clearly, for qubits it is enough to consider the Pauli-$Z$ operator for the choice of $A$ in Eq.~(\ref{PDC-condition}). Then, the PDC gate for an arbitrary choice of the operator $A$ can be obtained by a local change of basis for all qubits. Thus, we would like to describe the $n$-body gates $U^{(n)}$ that satisfy the PDC condition in the form
\begin{equation}\label{PDC-Z-gate}
     [ U^{(n)} , Z_{i} - Z_{j} ] = 0 \quad \forall i,j \in \{1,\ldots,n\},
\end{equation}
where $Z_{i}$ is the Pauli-$Z$ operator that acts on the site $i$. 

% We constrain the gate to act in such a way that for any pair of qubits, it commutes with the corresponding staggered-magnetization:
% \begin{equation}\label{PDC-gate}
%      [ U^{(n)} , Z_{i} - Z_{j} ] = 0 \quad \forall i,j \in \overline{1,n},
% \end{equation}

% where $Z_{i}$ is the  Pauli-$Z$ operator acting on site ${i}$. In what follows, we will call such gates Staggered-Magnetization Conserving (PDC). As discussed in the next subsection, this property is crucial for the construction of local integrals of motion in circuits made from such gates. Before proceeding in this direction, %let us show 
% we parametrize the most general form of gates that satisfy the condition~\eqref{eq:PDC_gate}. 

The family of pairwise difference operators $Z_i - Z_j$ is diagonal in the computational basis $\ket{\mathbf{s}}=
\ket{s_1,\ldots s_n}$. Moreover, two basis states give the same eigenvalues for every $Z_i-Z_j$ iff either % for every $D_{ij}$ and has a unique two-dimensional kernel: the uniform states 
$\ket{\mathbf{s}}=\ket{0^n}$ or $\ket{\mathbf{s}}=\ket{1^n}$ (both give all zeros).
% \begin{equation}
%     (Z_i - Z_j) \ket{0^n} = (Z_i - Z_j)\ket{1^n} = 0.
% \end{equation}
Therefore, %it is instructive to see how the gate 
the quantum gate $U^{(n)}$ satisfying~(\ref{PDC-Z-gate}) must act as a unitary $u=\begin{pmatrix} 
\alpha & \beta \\
\gamma & \delta
\end{pmatrix} \in U(2)$ in span$\{\ket{0^n},\ket{1^n}\}$:
\begin{equation}
\begin{aligned}
    &\ket{0^n} \text{ } \xrightarrow{U^{(n)}} \text{ } \alpha \ket{0^n} + \beta \ket{1^n}, \\
    &\ket{1^n} \text{ } \xrightarrow{U^{(n)}} \text{ } \gamma \ket{0^n} + \delta \ket{1^n},
\end{aligned}
\end{equation}
and be purely diagonal in the remaining space. As such, the most general $n$-body gate $U^{(n)}$ satisfyingthe condition~(\ref{PDC-Z-gate}) has the matrix form
\begin{equation}
    U^{(n)}  = 
\begin{pmatrix}
\alpha & & &  &\beta \\
  & e^{i\varphi_1} & & \\
 &  & \ddots & \\
 & &  & e^{i \varphi_{2^{n}-2}} & \\
\gamma  & & & & \delta 
\end{pmatrix}, \label{eq:PDC_gate}
\end{equation}
and can be parametrized by $2^{n}+2$ real parameters.

With the help of creation and annihilation operators $\sigma^{+}_j = \frac{1}{2}(X_{j} +i Y_j)$, $\sigma^{-}_{j} = \frac{1}{2}(X_{j} - i Y_{j})$ the PDC gate can also be written in exponential form as
\begin{equation}
    U^{(n)} = \exp\left(i \sum_{S \subseteq [n]} h_S \prod_{j \in S}Z_j \right)   \exp \left( i [J \prod_{j=1}^{n} \sigma_{j}^{+} + h.c.] \right) 
\end{equation}
% \begin{equation}\label{PDC-gate-exp}
%     U^{(n)} = e^{i \phi} \exp \left(i \sum_{j=1}^{n} h_{j} Z_j \right) \exp \left( i [J \prod_{j=1}^{n} \sigma_{j}^{+} + h.c.] \right) 
% \end{equation}
where in the first exponent the sum is over all possible subsets of $[n] = \{1, ..., n\}$ with the convention $\prod_{j \in \emptyset} Z_j \rightarrow 1$. The constants $h_S \in \mathbb{R}$ and $J \in \mathbb{C}$ can be chosen arbitrarily (note that the total amount of real variables in this parametrization is also $2^{n}+2$). %In particular, 
From this expression, we see that one can also consider Hamiltonian densities of the form
\begin{equation}\label{PDC-hamiltonian}
    H^{(n)} = J \prod_{j=1}^{n} \sigma_{j}^{+} + h.c. +\sum_{S \subseteq [n]} h_S \prod_{j \in S}Z_j,
\end{equation}
which satisfy the PDC condition $[ H^{(n)} , Z_{i } - Z_j  ] = 0$. The construction of local charges in Section~\ref{construction-charges} applies for such densities as well. 

Finally, we emphasize that the construction of local charges and the main results of this paper do not require fine-tuning of the PDC gate parameters. The primary role of these parameters is to control the classical limit.

% Lastly, as already prefaced, the construction of local charges and the main results of this paper do not require fine-tuning of the parameters of the PDC gate. However, the knob for which it may be useful to tune them is to access the classical limit.

% \cj{To be consistent with the number of parameters for the unitary gate ($2^n+2)$ we should really be including all diagonal Z-strings of any degree, so $\sum_{S \subset [n]}  K_S \prod_{j\in S}Z_j$}.

% At this point, we would like to make a few remarks. Firstly, we consider here the staggered magnetization instead of the usual one because the condition $ [Z_{i} + Z_{j} , U^{(n)}] = 0$ $\forall i,j \in \overline{1,n}$ forces the gate to act trivially already for the case of $n=3$. Secondly, instead of Pauli-$Z$ operators, one can consider gates $U^{(n)}$ that commute with $A_{i}-A_{j}$, where $A$ is an arbitrary non-degenerate one-qubit operator. But of course, such gates are reduced to PDC gates~\eqref{eq:PDC_gate} by a local change of basis on each qubit. And lastly, while in this letter we will be concentrated on the case of qubits, our construction is straightforwardly generalized for qudits with any local dimension. 

\subsection{Classical limit}
There are two ways to make PDC gates classical. Firstly, in the case of qubits, one can choose the $U(2)$ matrix in~\eqref{eq:PDC_gate} to be 
\begin{align}\label{CA}
   u_{CA} = \begin{pmatrix}
    0 & 1 \\
    1 & 0
\end{pmatrix},
\end{align}
which is just a permutation operator. Together with zero phases $\varphi_{j}=0, \> \forall j$, the gate~\eqref{eq:PDC_gate} maps product states to product states. Thus, this gate is purely classical, and the circuits constructed from such gates describe the dynamics of a 2-state reversible cellular automaton (CA). This was studied in \cite{kasim2024deterministicmanybodydynamicsmultifractal} and is a particular case of what has been termed parity-check automata.

Another possibility to make the gate classical, but not with a discrete phase-space, is to considerclassical spins -- tops $\{ \boldsymbol{S}_{j}\in\mathbb R^3 \}_{j=1}^{n}$, i.e. continuous degrees of freedom that take values on the sphere $\boldsymbol{S}_j^2 =1$ and satisfy the Poisson algebra
\begin{equation}\label{class-spins}
    \{ S_{j}^{\alpha} , S_{k}^{\beta} \} = \delta_{jk} \varepsilon_{\alpha\beta\gamma} S_{j}^{\gamma}
\end{equation}
In the case of the classical Hamiltonian density $h^{(n)}_{cl}$ we should impose the Poisson-commutativity constraint
\begin{equation}
    \{ H_{cl}^{(n)} , S_{i}^z - S_{j}^{z} \} =0 \quad \forall i,j \in \overline{1,n}.
\end{equation}
Representing classical spins using spherical coordinates $\boldsymbol{S}_j = (\cos \varphi_j \sin \theta_j , \text{ } \sin \varphi_j \sin \theta_j , \text{ } \cos \theta_j  )$, this constraint simply reads as $ \left( \partial_{\varphi_i} - \partial_{\varphi_j}\right) H_{cl}^{(n)} =0 $. So, the Hamiltonian density can depend only on the sum of all $\varphi_{j}$:
\begin{equation}
    H_{cl}^{(n)} = H_{cl}^{(n)} (\{ \cos \theta_{j}\}_{j=1}^{n} , \sum_{j=1}^{n} \varphi_{j}).
\end{equation}
The simplest non-trivial example is just $H_{cl}^{(n)} = J \prod_{j=1}^{n} S_{j}^{+} + c.c. = 2J \cos\left(\sum_{j=1}^{n} \varphi_{j} \right) \prod_{j=1}^{n} \sin \theta_j,$ which is similar to the qubit Hamiltonian density~(\ref{PDC-hamiltonian}).

% Such cellular automata on a square and hexagonal lattices have been extensively studied in~\cite{kasim2024deterministicmanybodydynamicsmultifractal}. For instance, the multifractal response of the dynamical correlation function has been found.

\subsection{PDC Circuits and Loop Charges}\label{construction-charges}

\begin{figure}[H]
  \centering
  \includegraphics[scale=0.7]{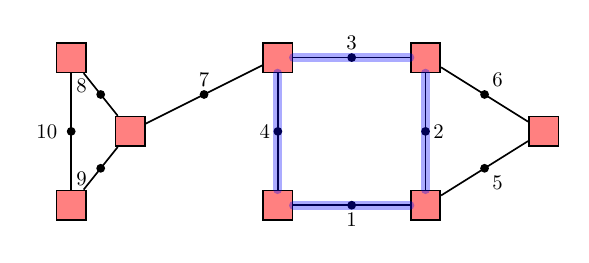}
  \caption{Example of a Circuit constructed on a graph. Vertices of a graph illustrated as red squares correspond to PDC gates of the form~\eqref{eq:PDC_gate}, while qubits (black circles) live on the edges. With the blue loop $\gamma = (1,2,3,4)$ one can associate a charge $M_{\gamma}$, see Eq.(\ref{charge-path}), that is conserved. }

    \label{fig:circuit-graph}
    
\end{figure}

Consider an arbitrary simple graph $G= (V,E)$ with a set of vertices $V$ and a set of edges $E$. To each edge $e \in E$ of the graph, we assign a dynamical degree of freedom, one example being a qubit. If the vertex $v \in V$ has a coordination number $n(v)$, we associate an $n(v)$-qubit PDC type gate~\eqref{eq:PDC_gate} to this vertex, which acts on all incident qubits. We restrict attention to graphs with minimum degree $\min_{v\in V} n(v)>1$, so that any fragmentation of the state space reflects symmetries or kinematic constraints rather than artifacts of leaf (degree-one) vertices. % (we will assume for simplicity that $\forall v$ $n(v)>1$ in order not to have isolated vertices). 
See Fig.~\ref{fig:circuit-graph} for an example.

We can use this graph construction to define Floquet circuits. The single-period Floquet operator, $\mathcal{U}_F$, is defined by applying the local gates in a particular sequence. The PDC gates can be chosen identically at every vertex for a homogeneous circuit, or drawn independently at random on each site.

Now, to each closed even-length loop $\gamma$ on a graph $G$ \footnote{More generally, for any closed walk, since vertices and edges may repeat.} %, the length of which $|\gamma|$ is an even number, 
one can associate a local operator that is conserved under the Floquet dynamics. Labeling the consecutive edges of the loop as $(e_1 , ... , e_{|\gamma|})$ and letting $Z_{j}$ act on the degree of freedom attached to each edge $e_j$, %with the same enumeration of degrees of freedom that belong to these edges, one can 
we define the associated charge as
\begin{equation}\label{charge-path}
    M_{\gamma} = \sum_{j=1}^{|\gamma|} (-1)^{j-1} Z_{j} \quad \forall \gamma \text{ s.t. } |\gamma| \text{ is even}.
\end{equation}
The even-length requirement comes from the fact that the first and the last qubits in the loop should have different expectation values of $Z_j$ in order to commute with the local PDC gate. For instance, in Fig.\ref{fig:circuit-graph} the even loop $\gamma=(1,2,3,4)$ yields $M_\gamma=Z_1-Z_2+Z_3-Z_4$, whereas odd loops such as $(1,5,4)$ or $(2,3,5)$ admit no consistent alternating assignment and thus no charge. Using that the gates of the circuit are of PDC type~\eqref{eq:PDC_gate}, one can easily verify that $ [ M_{\gamma} , \mathcal{U}_F ] = 0$. Thus, each even-length closed loop $\gamma$ defines a local symmetry of the system. 

%The condition for an even length of the loop comes from the fact that the first and the last qubits in the loop should have different signs of $Z$ operators to commute with the local gate. For example in Fig.~\ref{fig:circuit-graph}, the even-length closed loop $\gamma = (1,2,3,4)$ defines a charge $M_{\gamma} = Z_{1} - Z_{2} + Z_{3} - Z_{4} $, while it is not possible to associate a charge with odd-length loops like $(1,5,4)$ or $(2,3,5)$.

\begin{figure}[H]
  \centering
  \includegraphics{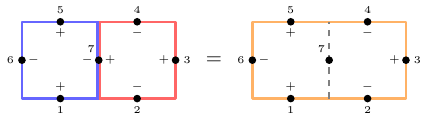}
  \caption{ The sum of charges $M_{\gamma_1}$+$M_{\gamma_2}$ that are defined on the blue $\gamma_1 = (1,7,5,6)$ and the red $\gamma_2 = (3,4,7,2)$ loops is the same as $M_{\gamma_3}$ for a merged orange loop $\gamma_3 = \gamma_1 + \gamma_2$.  Pluses and minuses indicate with which signs each spin contributes to the charge, Eq.(\ref{charge-path}).}

    \label{fig:two-charges}
    
\end{figure}

Now, even if we can define a charge~\eqref{charge-path} for any loop $\gamma$ on a graph, not all of them are linearly independent. If two loops $\gamma_1$ and $\gamma_2$ share the same edge $e$ and the corresponding charges have different signs on it in Eq.(\ref{charge-path}), the sum of these charges will be the same as a charge that is defined on a loop $\gamma_1 + \gamma_2$ obtained by merging   $\gamma_1$ and $\gamma_2$ along a common edge, see Fig.~\ref{fig:two-charges}.

Thus, when working with a PDC circuit on a graph $G$, it is important first to fix a basis of loops in such a way that for any loop $\gamma$ on the graph, the charge $M_{\gamma}$ could be expressed as a linear combination of basis charges. From a graph theory perspective, this correspond to fixing an even-length cycle basis on a graph $G$ and to associate a charge to each element of this basis. For a bipartite graph $G$, all loops have even length, and the even-length cycle bases coincide with the usual cycle bases.

More generally, one can derive under which conditions on the weights $w_{e}$ the operators of the type
\begin{equation}\label{weight-charge}
    M_{w} = \sum_{e \in E} w_{e} Z_{e}
\end{equation}
commute with the Floquet generator $\mathcal{U}_{F}$. As with loop charges (\ref{charge-path}), we will actually impose a much stronger constraint that the operator $M_{w}$ commutes separately with all local gates, not only with the whole $\mathcal{U}_F$. Thanks to the PDC condition (\ref{PDC-Z-gate}), the commutativity of $M_{w}$ with the local gate at an arbitrary vertex $v$ takes place if 
\begin{equation}\label{comm-cond}
    \sum_{e \in n(v)} w_{e} = 0 \quad \forall v.
\end{equation}
Defining the vertex-edge incidence matrix $B$ with matrix elements
\begin{equation}
B_{ve} =
\begin{cases}
1, & \text{if vertex $v$ is incident to edge $e$,} \\
0, & \text{otherwise,}
\end{cases}
\end{equation}
the condition (\ref{comm-cond}) can be written simply as $Bw = 0$. And thus, conserved quantities in the form (\ref{weight-charge}) are described by the kernel $\text{Ker}(B)$ of the incidence matrix. In particular, all loop charges of the form (\ref{charge-path}) belong to this kernel, and even-length cycle bases form a basis in this kernel as well. %For instance, if the graph $G$ is connected and bipartite, the number of independent charges is given by $\text{dim } \text{Ker}B = |E|-|V|+1$.

Finally, while the graph $G$ on which the circuit is defined can have any dimensionality, the conserved charges (\ref{charge-path}) are always defined on loops, which are 1-dimensional objects. Thus, these charges can be an example of 1-form symmetries~\cite{Gaiotto_2015,McGreevy_2023} of $U(1)$ type in discrete-space systems. It can be an interesting exercise to generalize our construction to higher-form charges that live on higher-dimensional surfaces.

\subsection{Examples of PDC circuits on regular graphs}\label{Examples}

\subsubsection{2D case: Square Geometry}

%\cj{On 2D lattices the natural local basis of Ker($B)$ is given by plaquette loops plus the minimal set of non-contractible loops for periodic boundary conditions. The former are maximally local integrals of motion, and the latter add a few global topopolgical charges. This coordinate choice is most convenient for several reasons: the sector labels are local (one charge number per plaquette), and the charges are local which is ideal for studying dynamics. Mention how from these basic loops one can fuse them.}

\begin{figure}[H]
  \centering
  \includegraphics[scale=0.95]{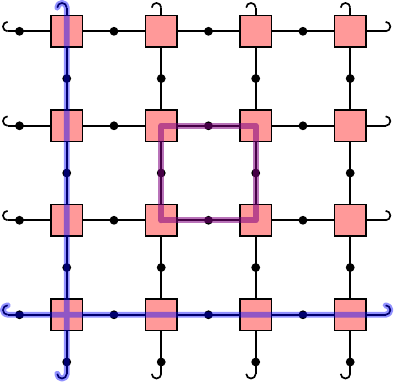}
  \caption{2D square $4 \times 4$ lattice constructed from  circuit PDC gates. We show a sample plaquette charge (purple) and both topological loops (blue). The full set of plaquettes together with the two non-contractible loops form a basis of loop symmetries (\ref{charge-path}). }
    \label{fig:2Dcase}
\end{figure}

The simplest example of a regular and translationally invariant graph with a 2-dimensional structure is a 2D square lattice, Fig.~\ref{fig:2Dcase} \footnote{Even though we assume here periodic boundary conditions, the choice of open boundaries does not affect the structure of the loop charges.%\cj{It does, we don't get the topological/transport loops.} \po{nono, you will still have it, in this case, boundary edges will have $n(v)=1$, so it is fine not to close the loop but to leave it open.} 
}. A lattice with $L_x \times L_y$ vertices (gates), has $2L_x L_y$ edges (degrees of freedom). We assume here both $L_x$ and $L_y$ to be even in order to have a bipartite lattice. In this case and for periodic boundaries, a natural basis of charges consists of local plaquette loops plus two non-contractible loops, giving a total of $L_x L_y +2$ independent charges: 

% \tp{I am not sure this formula is correct? Think of a small $2\times 2$ lattice with pbc. It has just $1+2$ independent loop charges. So maybe the correct expression is $(L_x-1)(L_y-1)+2$???} \cj{I think Lx Ly - 1 + 2}
\begin{align}
    M_{p(x,y)}&= Z_{x,y}-Z_{x-1,y}+Z_{x,y+1}-Z_{x+1,y}, \\ M_{\gamma_w} &= \sum_{j=1}^{L_{w}} (-1)^{j-1} Z_{j}, \quad w=x,y\,.
\end{align}
The plaquette charges are maximally local integrals of motion, while the winding loops furnish two global topological charges, see Fig.~\ref{fig:2Dcase}. In terms of integrability, at least for classical spins~(\ref{class-spins}), one can see that the amount of loop charges is not enough to guarantee the Liouville integrability of the system. %This coordinate choice is most convenient for several reasons: the sector labels are local (one charge number per plaquette), and the charges are local which is ideal for studying dynamics.
%The basis of charges is constructed from all possible elementary plaquettes of the lattice and two non-contractible loops that correspond to two non-trivial cycles of a torus, see Fig.~\ref{fig:2Dcase}. 
%The total number of charges %in this basis 
%is $L_xL_y+2$. 

Here, we note the similarity of our charges with the symmetries of the toric code proposed by Kitaev~\cite{Kitaev_2003}. In the toric code on a torus, the 4-fold degenerate ground-state subspace has the following symmetry. Any loop on a lattice or on its dual defines a symmetry, which is a product of Pauli $Z$ or Pauli $X$ operators across the corresponding loop. Therefore, such loop charges are $\mathbb{Z}_2$ in nature in contrast to our $U(1)$ charges~(\ref{charge-path}). Contractible loops act trivially on the ground states, while the four independent non-contractible loops (two on the lattice, two on the dual lattice) do not commute between themselves, and so, act nontrivially within this space. These loop operators serve as the logical $X$ and $Z$ operators for two encoded qubits -- a structure that realizes topological order and provides topological protection for quantum memory~\cite{Kitaev_2003,Nayak_2008}. However, in our case, the algebra of loop charges~(\ref{charge-path}) is strictly abelian, since all of them are constructed using only Pauli $Z$ operators. We can therefore protect at most one classical bit and detect only $X$-type errors. It could be worthwhile to explore generalizations of our construction that realize topological order.

In the case of spin-$1/2$ degrees of freedom, the plaquette charge $M_{p}$ takes on five distinct values: $\{-4, -2, 0 , 2 ,4\}$. The extremal values $\pm4$ correspond uniquely to the Néel configurations $0101$ and $1010$. Crucially, these plaquettes are completely frozen under the dynamics. Their immobility creates rigid barriers that split the system into disconnected dynamical regions. This is the microscopic origin of Hilbert space fragmentation.
%The values $\pm4$ correspond to Néel configurations on the loop: $0101$ and $1010$. %\cj{Maybe we make this generic: If the elementary plaquette has $n$ spins, $M_k$ can take on $n+1$ values. The minimal $n$ so that we have more than just the set $\{-4,0,+4\}$ in the $\Pi_k$=+1 parity sector is $n=8$. Would be important to realize systems where we have local areas which are not just fully frozen or maximally connected, similar to what Yusuf could do with the $\pm 2$ loops for the classical circuit.}. 
%Such configurations are frozen under the dynamics. 
By contrast, the intermediate values $\pm2$ (4 configurations each) and $0$ (6 configurations) admit nontrivial dynamics. One can therefore view the plaquette charge as a measure of how constrained a local region is: $M_p = \pm4$ being fully inert, while smaller charges allow nontrivial evolution. Consequently, the dynamical properties of an initial state depend sensitively on the distribution of plaquette charges it contains. The simplest macroscopic descriptors of such a distribution are the densities
%Hereby, the magnitude of the plaquette charge defines how 'dynamical' the degrees of freedom on that plaquette can be. One can expect that the dynamical properties of an initial state can highly depend on the distribution of the values of the plaquette charges in this state. The simplest example of macroscopic quantities that characterize the distribution of plaquette charges in a state is the densities of the plaquettes with the same values of the plaquette charge
\begin{equation}\label{densitites-nq}
    n_q = \frac{1}{L_x L_y} \{ \#\text{plaquettes} \quad \text{s.t.} \quad M_{p} = q \} .
\end{equation}
In the related work of our group~\cite{Yusuf}, the cellular automaton limit~(\ref{CA}) of the circuit on a square lattice is analyzed in greater detail. The densities~(\ref{densitites-nq}) are used there to define ensembles of initial states. Depending on the choice of these densities, the system exhibits either a localized or a delocalized phase with respect to information spreading, and these two phases are separated by a second-order phase transition. Although verifying this numerically for the quantum circuit is more challenging, we expect a similar transition to occur in the quantum case as well.

The topological charge $M_{\gamma_w}$ provides a global constraint on the dynamics: it can take on values in the range $-L, -L+2, \dots, 0, 2, \dots, L$, reflecting its role as an extensive conserved quantity. Since it is defined as a sum over staggered local densities along a non-contractible loop, this charge can be naturally interpreted as a measure of transport across the system. In this sense, it complements the local plaquette charges by encoding global flow constraints.

\subsubsection{1D case: Ladder Geometry}

The simplest graph in 1D can be constructed as follows 
\begin{figure}[H]
  \centering
  \includegraphics[scale=1.1]{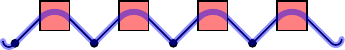}
  %\caption{\cj{Write caption} \po{not to put caption can be also fine for such a small figure}}
  %The circuit constructed from the PDC gates on a square lattice. Charges for all plaquette loops (violet) together with two non-contractible loops (blue) form a basis of loop symmetries (\ref{charge-path}). }
    \label{fig:1d-example}
    
\end{figure}
But of course, such choice of the graph lacks an extensive amount of loop symmetries. The only present loop in this case is going through the whole system (blue line), while the corresponding charge is just the global staggered-magnetization. Flipping even degrees of freedom $s_{2i} \rightarrow - s_{2i}$, the system becomes the most general circuit with nearest-neighbor 2-body interactions and conservation of magnetization. Interestingly, for the case of spin-$1/2$ degrees of freedom and homogeneous gates acting in a brickwall manner (for one Floquet period, firstly all odd gates are applied, after all even), the conservation of the global magnetization turns out to be enough to guarantee the Yang-Baxter integrability of the circuit~\cite{Znidaric_2025}.

The example of the 1D graph, where the loop charges are still present is the ladder geometry, where 3 rows of degrees of freedom interact by means of 3-body PDC gates as illustrated in Fig.~\ref{fig:ladder-geometry-2}. % \begin{figure}[H]
%   \centering
%   \includegraphics[scale=1.1]{images_main/ladder-geometry.pdf}
%   \caption{The 1D circuit constructed from the PDC gates with the ladder geometry. The basis of loop charges can be build from all elementary plaquette loops (violet) and one non-contractible loop (blue). }
%     \label{fig:ladder-example}
%\end{figure}
In this geometry, the basis of loop charges is essentially identical to the square lattice. In the case of the $3L$ degrees of freedom ($L$ in each row), there are $L$ charges that correspond to elementary plaquette loops, as well as 1 topological charge constructed from the non-contractible loop.

\section{Case Study of the Ladder Geometry}

\begin{figure}[]
\centering
\includegraphics[scale=1.1]{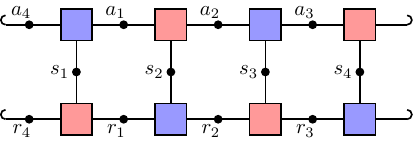}
\caption{Circuit with ladder geometry of length $L=4$.}
\label{fig:ladder-geometry-2}
\end{figure}

As the simplest model that still captures the relevant physics, the 1D ladder with spin 1/2 degrees of freedom will serve as our case study. We now introduce precise labels for its degrees of freedom (see Fig.~\ref{fig:ladder-geometry-2}). The Floquet protocol is applied in two steps, where we first apply the gates acting on the blue sublattice, followed by the gates acting on the red sublattice (brick wall protocol).
%\cj{Let's put the beginning of the next section also here, to conclude the set up. Then we start directly with studying HSF in the case study?} \po{Yes, you are right. Also, two pitures are a bit redundant. Maybe we can use the second one where blue-red gates with labels of bits are illustrated. Then indeed we can proceed here the definition of a Floquet.}
%\cj{Ok I will change this.}
% The exact LIOMs for this geometry are thus
% \begin{align}
%     M_k&= Z_{s_k}-Z_{a_k}+Z_{s_{k+1}}-Z_{r_k}, \\ M_{\rm cycle} &= \sum_{j=1}^L (-1)^{j-1} Z_{r_j}.
% \end{align}
%Here we go again. (describe floquet protocol, mention that spin-$1/2$)
We also note that there is an additional discrete symmetry in the system, which we resolve to reach larger system sizes for numerical simulations. Defining the $\mathbb{Z}_2$-parity operator of the $k$-th plaquette as
\begin{equation}
\Pi_{k} = Z^{(s)}_{k} Z^{(a)}_{k} Z^{(s)}_{k+1} Z^{(r)}_{k},
\end{equation}
we restrict ourselves to the parity-plus sector where $\Pi_k =1$ for all plaquettes. In the last equation, the superscript denotes on which type of qubits ($r$, $s$ or $a$) the Pauli-$Z$ operator acts, see Fig.~\ref{fig:ladder-geometry-2}.
%a particular sector of the circuit. 
%of the $k$-th plaquette in the ladder commutes with our circuit 
One can check that this parity operator commutes with the Floquet unitary simply by expressing it as $\Pi_{k} = e^{i\frac{\pi}{2}M_{k}}.$ Working inside the fully parity-plus sector, %where $\Pi_k =1$ for all plaquettes allows us to express all top qubits as 
we can express the qubits in the top row of Fig.~\ref{fig:ladder-geometry-2} as
\begin{equation}
    a_k = \left( s_k + s_{k+1} + r_k \right) \text{ mod }2.
\end{equation}
Thus, in this sector only qubits $\{ s_k\}_{k=1}^{L}$ and $ \{ r_{k} \}_{k=1}^{L}$ can be treated as dynamical degrees of freedom, effectively decreasing the number of qubits from $3L$ to $2L$. We label the states from this sector as $\ket{s_1 \text{ } r_1 \text{ }... \text{ }s_L \text{ } r_L}$ with $s_{i}, r_{i} \in \{ 0, 1 \}$, specifying the values of only the middle and bottom row. In the parity-plus sector, the plaquette charge can take on only three distinct values $M_p^+=\{-4,0,4\}$. The dynamics are governed by the distribution of these plaquettes: fully frozen ones with $M_p=\pm4$ act as rigid barriers, while $M_p=0$ plaquettes (6 configurations) remain dynamical. Their arrangement sets the pattern of barriers and active islands that fragments the Hilbert space, as we discuss next.
%the values of the plaquette charges $M_{k} = (s_k+s_{k+1})-r_k(1+s_k s_{k+1})$ can be only $\pm4$ or $0$. The values $\pm4$ correspond to frozen $0101$ and $1010$ configurations of the plaquettes. 

%\cj{Some intro sentence} \po{what if first few paragraphs from the previous sections will be here instead of intro sentence? These paragraphs suit perfectly} \cj{Yes sounds good!}

\subsection{Hilbert Space Fragmentation}

\begin{figure}[H]
  \centering
  \includegraphics[scale=0.8]{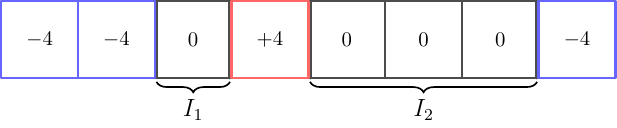}
  \caption{ Two dynamically decoupled islands, $I_1$ and $I_2$, separated by plaquettes with $M_k = \pm 4$.}
    \label{fig:islands}
\end{figure}

\begin{figure}[ ]
\centering
\includegraphics[width=0.9\columnwidth]{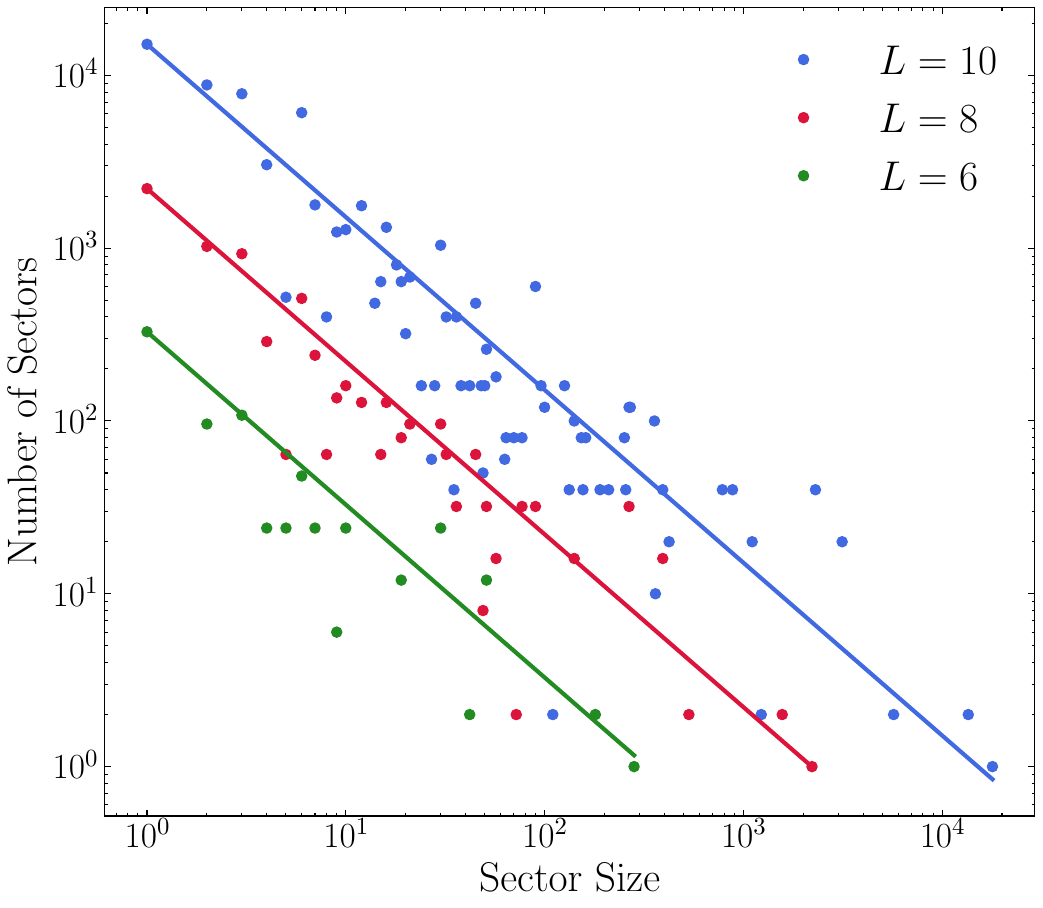}
\caption{Number of sectors in the $\Pi_k=+1,\> \forall k$ parity sector as a function of the sector size for three different system sizes of the ladder circuit. Straight lines of the form $1/x$ are plotted to guide the eye. A ladder with $L$ sites has $2L$ qubit degrees of freedom.}
\label{fig:sectors}
\end{figure}

Since all loop charges of type~(\ref{charge-path}) commute between themselves and are diagonal in the computational basis, their presence leads to the fragmentation of the full Hilbert space into dynamically decoupled sectors. The number of sectors of different size can be deduced from the matrix elements of the Floquet operator $\mathcal{U}_{F}$ in the computational basis, and is shown in Fig.~\ref{fig:sectors}. %The minimal fragmentation comes from the set of labels $\mathcal{M}=(\{M_k\},M_{\gamma})$. 
The minimal fragmentation comes from the set of plaquette charges $\mathcal{M}=\{M_k\}_{k=1}^{L}$: 
\begin{align}
\mathcal{H}&=\bigoplus_{\mathcal{M}}\mathcal{H}_{\mathcal M}.
\end{align}
Moreover, each $\mathcal{H}_{\mathcal{M}}$ is fragmented due to the presence of the cycle charge $M_{cycle} = \sum_{j=1}^{L}(-1)^{j} Z^{(r)}_{j}$. For a given choice of $\mathcal{M}$, the frozen plaquettes with $M_k=\pm4$ split the ladder into dynamically decoupled ''active'' islands $I_1,... , I_c$, all of which have $M_k =0$, see Fig.~\ref{fig:islands}. 

On such islands we can define $C_{I} = \sum_{j\in {I}} (-1)^{j-1}Z^{(r)}_{j}$. Since the Floquet dynamics preserve $\sum_I C_I$, and no gate can act on more than one island, we conclude that in $\mathcal{H}_{\mathcal{M}}$ each $C_I$ is independently conserved. Thus, the space $\mathcal{H}_{\mathcal{M}}$ can be further decomposed into fragments characterized by different values of $C_{I}$:
\begin{equation}
    \mathcal{H}_{\mathcal{M}} = \bigoplus_{\mathcal{C}}\mathcal{H}_{\mathcal{M},\mathcal{C}} \quad \text{with}\quad \mathcal{C} = \{ C_I \}_{I \in \mathcal{M}}. 
\end{equation}
Naively, the island $I$ constructed from $|I|$ plaquettes can take on $|I|+1$ different values of $C_{I}$. This would suggest that the total number of sectors in $\mathcal{H}_{M}$ is $\prod_{I \in \mathcal{M}} (|I|+1) $. However, the possible values of $C_I$ are slightly restricted by the boundary plaquettes on the left and the right of the island. For instance, if the island with odd length $|I|$ is surrounded by $+4$ plaquettes on both sides, it can't be in the configuration $'1010...1'$ that corresponds to the lowest value of $C_I = -|I|$. Similarly, if an even-length island is surrounded by $-4$ on the left and $+4$ on the right, it can't be in the configuration $'0101...01'$. Numerically we find that the total number of sectors for $2L$ effective spins is $3^{L}+L$, a form so elegant it strongly suggests an underlying combinatorial structure, but we have not been able to find it yet. 
% (I see numerically that putting all these conditions, the number of sectors is like that. Maybe it is worth thinking about how to compute it analytically). %\cj{Yes the $3^L+L$ is what led me to realize the Hilbert space was further fragmented. Here was the previous comment:}%\cj{One thing confuses me regarding the number of Krylov sectors: In the +1 parity sector, we find numerically that this number is exactly $3^L+L$, so asymptotically scales as $3^L$. But that's what we'd expect from the following naive argument: We have $L$ plaquette values $M_k$ to assign, which can take on the sign values $\{-1,0,+1\}$ (corresponding to $M_k$ values $\{-4,0,+4\}$), which gives $3^L$ labels.}

%\cj{But separately, we know that $M_{k} = \pm 1$ can never be adjoint (the frozen states $1010$ and $0101$ can never exist next to each other). Phrased this way, the number of labels is given by solving the transfer matrix problem given by the matrix}
% \begin{align}
%     \begin{pmatrix}
%         1&1 &0 \\ 1&1&1\\0&1&1
%     \end{pmatrix}.
% \end{align}
% \cj{It is written in the basis $\{-1,0,+1\}$, and has eigenvalues $1, 1\pm \sqrt{2}$, so the predicted number of sector goes as $1+(1-\sqrt{2})^L + (1+\sqrt{2})^L$, which is slightly smaller than $3^L$. So why do our numerics predict $3^L+L$ for the total number of Krylov sectors? Is there a difference between the \textbf{label} $(\{M_k\},M_{\gamma})$ and the \textbf{Krylov} sectors?}

The dimension $\mathcal{D}_{\text{max}}$ of the maximal sector follows by counting configurations with all plaquette charges equal to zero. 
%The fact that the maximal sector dimension $\text{dim} \mathcal{H}_{\text{max}}$ grows as $3^{L}$ can be understood as follows. One can compute the number $N_{\text{pl}=0}$ of such configurations where all the plaquette charges are equal to $0$. 
These configurations are characterized by the absence of $'101'$ and $'010'$ patterns for all $(r_{i}, s_{i+1} , r_{i+1})$ with $i \in \{1,\ldots L\}$. Fixing the bottom row $\boldsymbol{r} = (r_1, ..., r_{L})$ first, we can deduce which values of $s_{i}$ are allowed. If $r_{i} = r_{i+1}$, then $s_{i}$ must be the same, i.e. $s_i = r_i = r_{i+1}$, whereas if $r_{i} \neq r_{i+1}$, then $s_{i}$ can take any value (zero or one). This can be encoded by the transfer matrix 
%$A=((1,2),(2,1))$,
 $A = \begin{pmatrix}
         1&2\\2&1
     \end{pmatrix}
 $, 
with entries $A_{r_{i} r_{i+1}}$. For a full row configuration $\boldsymbol{r}$ on $L$ even plaquettes, we take $\prod_{i=1}^{L} A_{r_{i} r_{i+1}}$, giving a total number of  %Thus, for a given row configuration $\boldsymbol{r}$, the number of allowed values of $s_i$ is encoded by the matrix element $A_{r_{i} r_{i+1}}$. While the total amount of allowed configurations with the down row $\boldsymbol{r}$ can be written as a product $\prod_{i=1}^{L} A_{r_{i} r_{i+1}}$. 
%Thereby, the total number of configurations with zero value of the plaquette charges $N_{\text{pl}=0}$ can be written as
\begin{equation}
    N_{\mathbf{p}=0} = \sum_{\boldsymbol{r}} \prod_{i=1}^{L} A_{r_i r_{i+1}} = \text{tr } A^{L} = 3^{L} +1.
\end{equation}
%In the last equality, we used that $L$ is even in our case. 
Importantly, we still need to resolve the topological charge $M_{\text{cycle}}$ to fully specify the sector. 
%In addition, all these states are fragmented by the value of the cycle charge $M_{\text{cycle}} = \sum_{j=1}^{L} (-1)^{j-1} Z_{r_j}$. 
This, however, does not change the asymptotic scaling $\mathcal{D}_{\text{max}} \sim 3^{L}$. We have thus shown that the fragmentation is strong, as measured by the ratio of dimensions of the maximal sector and the full Hilbert space going to zero in the thermodynamic limit. 

% We have checked numerically that each dynamical sector in the computational basis can be labeled by unique values of charges~(\ref{charges-ladder}), meaning that the fragmentation is governed only by constructed symmetries. The total number of sectors and the dimension of the maximal sector (that correspond to the $0$ values of all charges) grow exponentially with the system size as $3^{L}$. This indicates that the fragmentation is strong, meaning that the ratio between the dimensions of the maximal sector and the full Hilbert space is equal to zero in the thermodynamic limit. 

\subsection{Dynamics} 

We now turn to the dynamical properties of the model. Our focus will be on two key probes of ergodicity breaking: the two-point correlation function of local observables and the return probability of the initial bitstring configuration. 
%\cj{In numerics we fix a representative $u\in U(2)$ away from the classical limit, but the qualitative dynamical behavior we report is robust across broad parameter choices.}
% We compare the values of these two quantities averaged over: 1) the complete basis of bitstring states or 2) the bitstring states from the maximal sector $\mathcal{H}_{0,0}$. 

As a local observable, we consider the %$ $Z_{s_k}$ (or $Z_{r_k}$) (actually $Z_{r_k}$ is a 
local operator $Z^{(r)}_{k}$ representing the charge density for loop charge $M_{cycle}$.%, maybe it is more interesting to take it as it is related to transport, right?) since this operator does not act between different sectors. 
We compare the infinite-temperature correlation functions 
\begin{equation}
\begin{aligned}
    &C(t) = \frac{1}{\mathcal{D}_{\mathcal{H}}} \text{Tr}_{\mathcal{H}} \left[ Z^{(r)}_{k}(t)Z^{(r)}_{k} (0) \right], \\
    C_{\rm max}&(t) = \frac{1}{\mathcal{D}_{ \text{max} } } \text{Tr}_{\mathcal{H}_{\text{max}}} \left[ Z^{(r)}_{k}(t) Z^{(r)}_{k} (0) \right]
\end{aligned}
\end{equation}
averaged over the whole space $\mathcal{H}$ and over the maximal sector $\mathcal{H}_{\text{max}}$ only, each of dimension $\mathcal{D}_{\mathcal{H}}$ and $\mathcal{D}_{\text{max}}$ respectively. As expected, due to the exponential amount of small sectors, the infinite-temperature correlation function $C(t)$ with time decays to a finite value even in the thermodynamic limit. This is known in the literature as ``operator localization''~\cite{Pai_2019,Sala_2020,Rakovszky_2020}. On the contrary, the restricted correlator $C_{\text{max}}(t)$ shows decay to a zero value (see Fig.~\ref{fig:ReturnProb} a)).

Next, we consider the dynamics of return probabilities%$ bitstring states $\ket{\boldsymbol{b}}$ from the whole basis and from the maximal sector only
\begin{equation}
\begin{aligned}
    P(t) &= \frac{1}{ \mathcal{D}_{\mathcal{H}}}  \sum_{s\in \mathcal{M}} \sum_{\ket{\boldsymbol{b}} \in \mathcal{H}_s } | \bra{\boldsymbol{b}} \mathcal{U}^{t}  \ket{\boldsymbol{b}} |^2 , \\
    P_{\text{max}}(t) &= \frac{1}{ \mathcal{D}_{\text{max} } } \sum_{\ket{\boldsymbol{b}} \in \mathcal{H}_{\text{max}} } | \bra{\boldsymbol{b}} \mathcal{U}^{t}  \ket{\boldsymbol{b}} |^2, \label{eq:ReturnProb}
\end{aligned}
\end{equation}
where we once more contrast the behavior when averaged over the full Hilbert space versus only the maximal sector. %\cj{Move to next section: Assuming each sector is quantum chaotic, the long-time average of a single state's return is its IPR in the Floquet basis. Substituting this into the expressions above, we find that $\overline{P} = N_{\rm sec}/\text{dim} \mathcal{H}$, $\overline{P}_{\rm max} = 1/D_{\mathcal{H}_{\rm max}} $ and $\overline{P_0} = 1/\text{dim} \mathcal{H}_0$. Since $\text{dim} \mathcal{H}\leq N_{\rm sec}D_{\mathcal{H}_{\rm max}}$, we have $\overline{P}\geq \overline{P_{\rm max}}$, consistent with the findings in \ref{fig:ReturnProb}}
%\cj{Assuming each sector is quantum chaotic (eigenvectors are random within $\mathcal{H}_s$, and no degeneracies. Latter is not true, but this will at worst raise the long-time plateau. let's go ahead for now), the long-time average of a single state's return is its IPR in the Floquet basis. Substituting this into the expressions above, we find that $\overline{P} = N_{\rm sec}/\text{dim} \mathcal{H}$, $\overline{P}_{\rm max} = 1/D_{\mathcal{H}_{\rm max}} $ and $\overline{P_0} = 1/\text{dim} \mathcal{H}_0$. Since $\text{dim} \mathcal{H}\leq N_{\rm sec}D_{\mathcal{H}_{\rm max}}$, we have $\overline{P}\geq \overline{P_{\rm max}}$, consistent with the findings in \ref{fig:ReturnProb}}. 
%As one can see from Fig.~\ref{fig:ReturnProb}, 
Fig. \ref{fig:ReturnProb} shows how the return probability in the full space $P(t)$ decays to its long-time (finite) value in a few time steps, while $P_{\text{max}}(t)$ decays over a much wider time window, a behavior common for ergodic dynamics~\cite{santos,Torres_Herrera_2018,Hopjan_2023}. After this decay, $P_{\text{max}}(t)$ also saturates, but to a much smaller plateau as compared to $P(t)$. Despite the maximal sector being ergodic as a whole, there exist bitstrings in it with persistent revivals, a phenomenon reminiscent of quantum many-body scars \cite{Turner_2018,Serbyn_2021}. Examples of such states are \textit{all}-$0$ and \textit{all}-$1$ bitstrings that belong to $\mathcal{H}_{\rm max}$. The dynamics of $P_0(t) = |\langle 0 | \mathcal{U}^{t} |0 \rangle |^2$ is also illustrated in Fig.~\ref{fig:ReturnProb}. 

In the long-time limit, the (average) return probability of a single state equals its IPR in the Floquet eigenbasis; this motivates analyzing IPR as an eigenstate diagnostic.

\subsection{Eigenstate diagnostics I: IPR}

\begin{figure}[ ]
\centering
\includegraphics[width=.45\columnwidth]{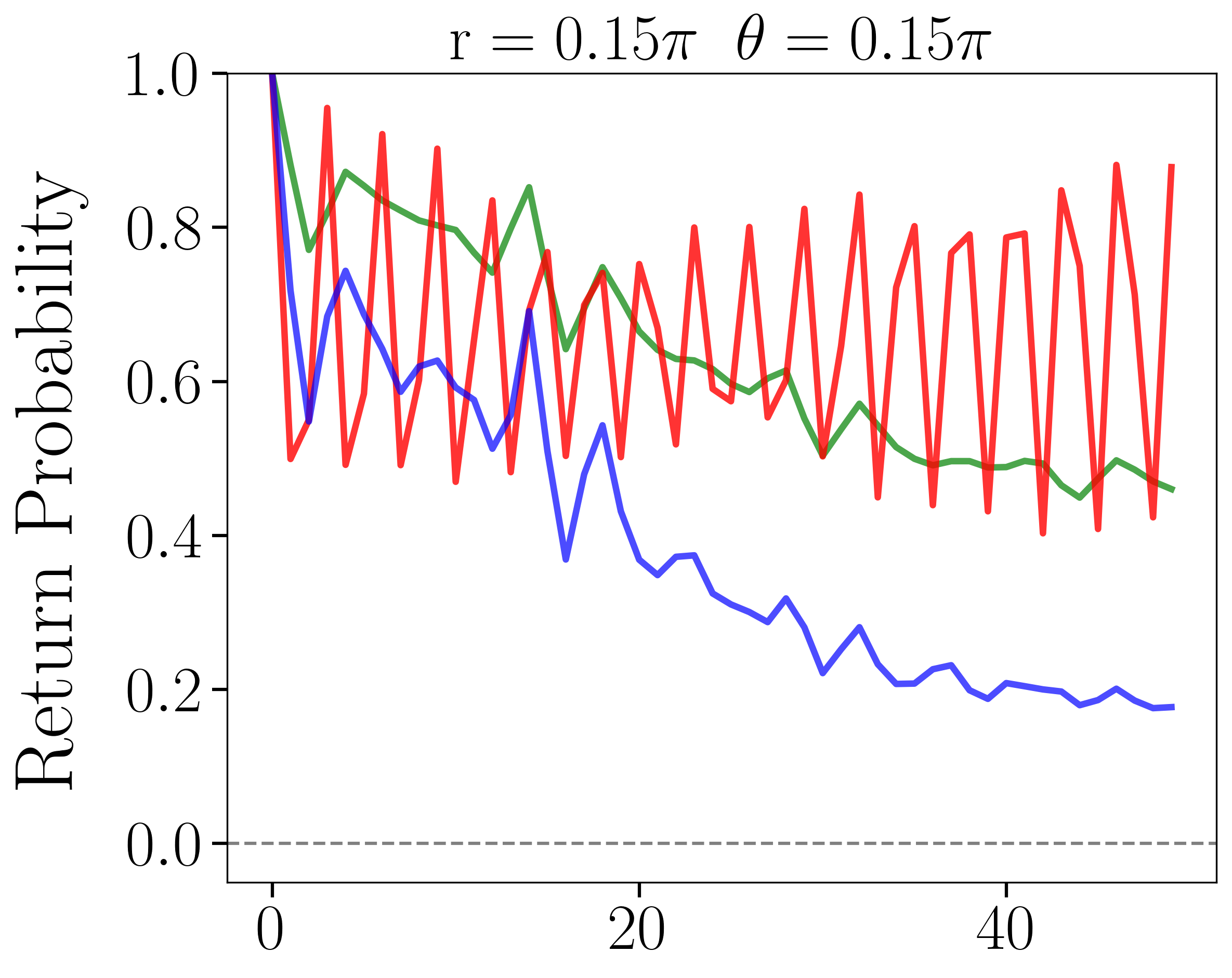}
\includegraphics[width=.45\columnwidth]{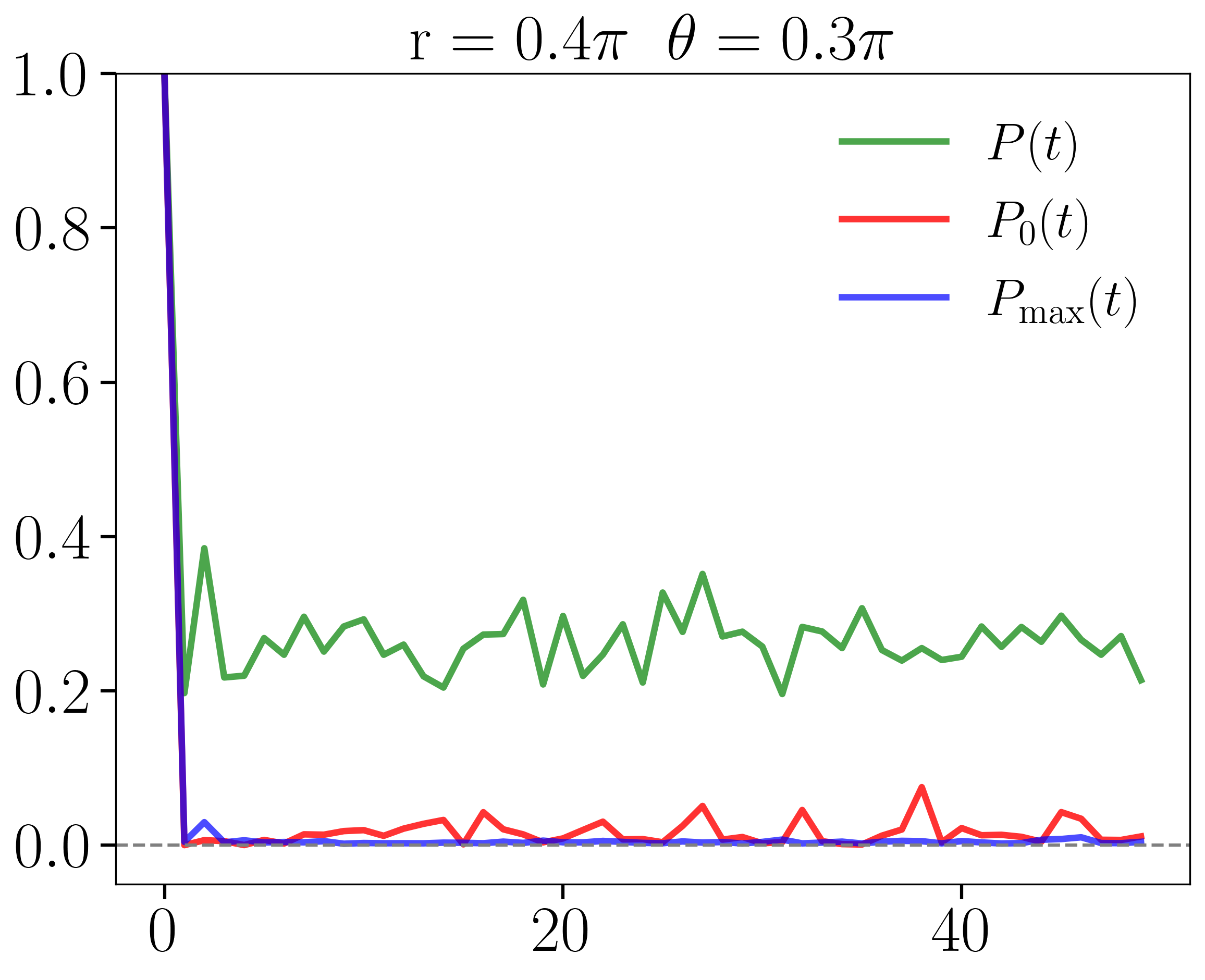}\\
\includegraphics[width=.45\columnwidth]{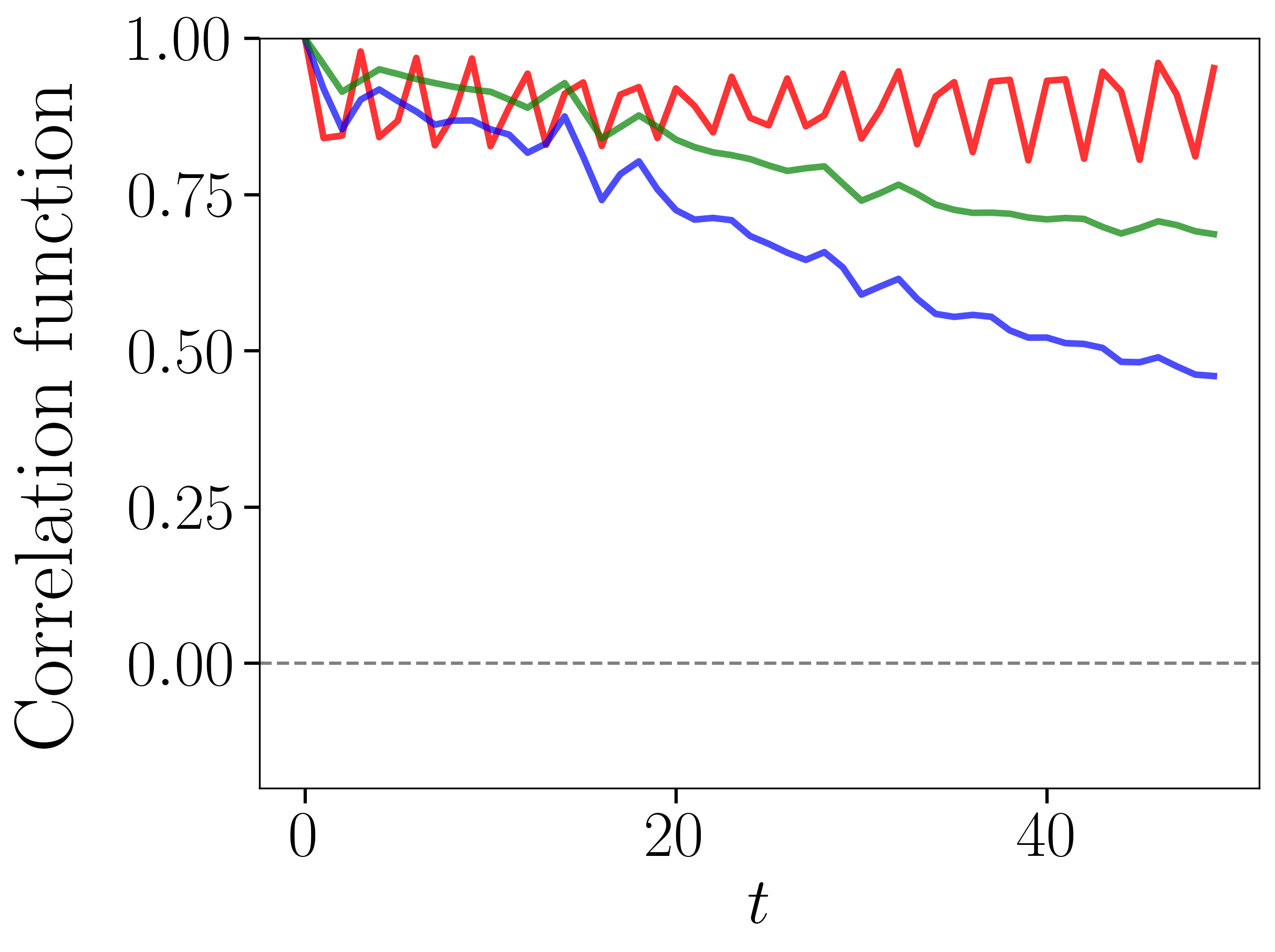}
\includegraphics[width=.45\columnwidth]{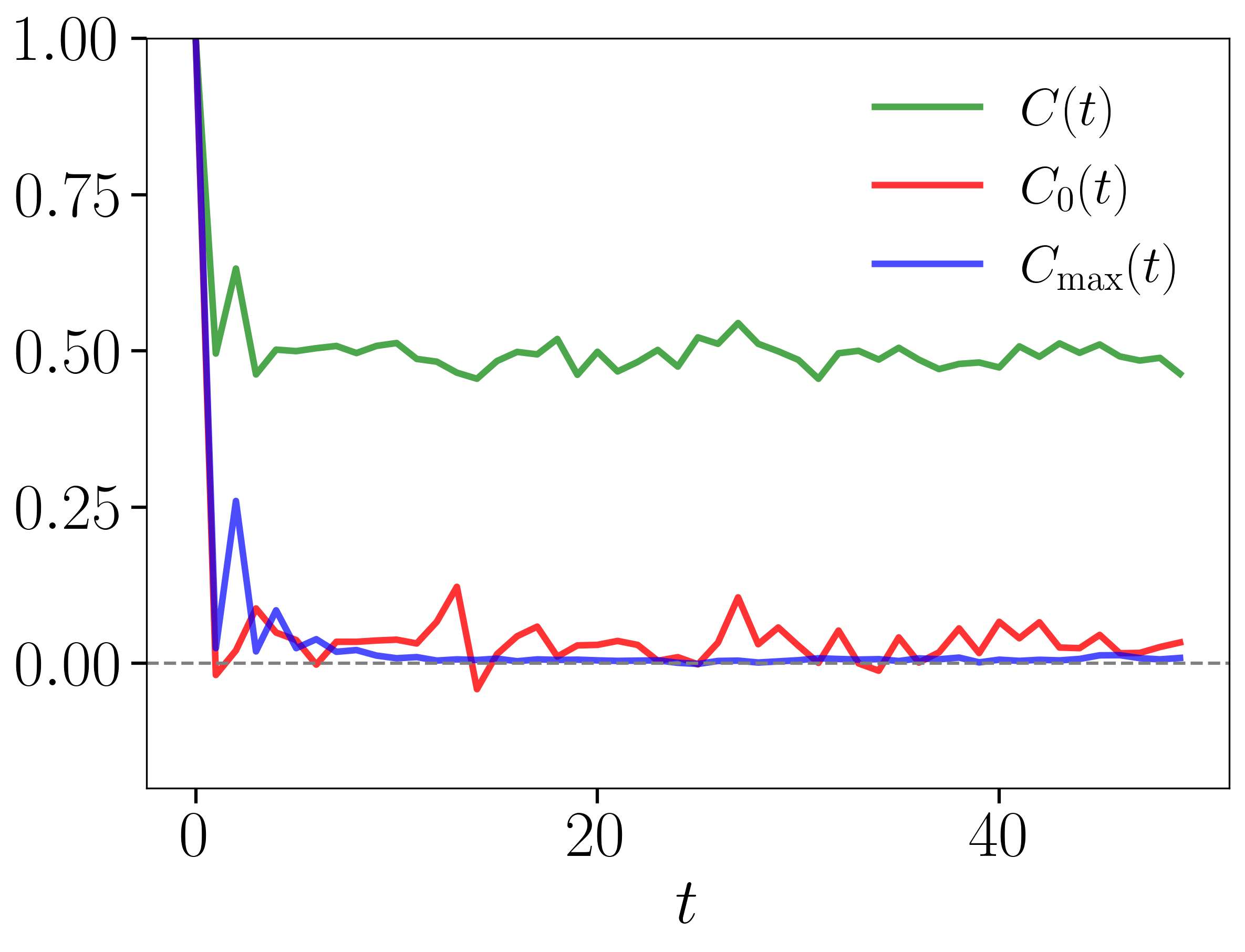}\\
\caption{Dynamics of the return probability $P(t)$ (top) and the local correlation $C(t)$ (bottom) for the PDC ladder. Curves shown: averages over all basis states $(P(t),\,C(t))$, the maximal sector $(P_{\mathrm{max}}(t),\,C_{\mathrm{max}}(t))$, and the all-zero state $(P_{0}(t),\,C_{0}(t))$. System sizes: $L=6$ for full-Hilbert-space averages and $L=8$ for the sector analysis. The unitary matrix for the PDC gate is $u=e^{\,i r(\sin\theta\,\sigma^{x}+\cos\theta\,\sigma^{z})}$ with $\theta=0.3\pi$. Left: less-ergodic regime $r=0.15\pi$, $\theta=0.15 \pi$. Right: more-ergodic regime $r=0.4\pi$, $\theta = 0.3 \pi$.}
%\tp{Can we show data for at least two different $L$ here?} \cj{This just added a lot of noise to the plot, but can easily re-instate.}}
\label{fig:ReturnProb}
\end{figure}

% \cj{Move to next section: Assuming each sector is quantum chaotic, the long-time average of a single state's return is its IPR in the Floquet basis. Substituting this into the expressions above, we find that $\overline{P} = N_{\rm sec}/\text{dim} \mathcal{H}$, $\overline{P}_{\rm max} = 1/D_{\mathcal{H}_{\rm max}} $ and $\overline{P_0} = 1/\text{dim} \mathcal{H}_0$. Since $\text{dim} \mathcal{H}\leq N_{\rm sec}D_{\mathcal{H}_{\rm max}}$, we have $\overline{P}\geq \overline{P_{\rm max}}$, consistent with the findings in \ref{fig:ReturnProb}}
The long-time saturation value, or more precisely, the time average of the return probability of a bitstring $\ket{\boldsymbol{b}} \in \mathcal{H}_{s}$ can be characterized by the Inverse Participation Ratio (IPR)
\begin{equation}
    \text{IPR}_{\boldsymbol{b}} = \sum_{n=1}^{\mathcal{D}_{\mathcal{H}_{s}}} | \langle \boldsymbol{b} | \varphi_{n} \rangle |^4,
\end{equation}
where $\{ \ket{\varphi_{n}} \}_{n=1}^{\mathcal{D}_{\mathcal{H}_s}}$ is a set of eigenvectors of the Floquet operator $\mathcal{U}_F$ in the chosen sector with eigenvalues $\varphi_n$. For bitstrings that are delocalized over the basis of eigenstates, one can expect the overlaps to behave as $|\langle \boldsymbol{b} | \varphi_{n}  \rangle|^2 \sim \mathcal{D}_{\text{max}}^{-1}$. As a result, the IPR for such states should scale as $\text{IPR}_{\boldsymbol{b}} \sim \mathcal{D}_{\text{max}}^{-1} $. This predicts the long-time saturation values of \eqref{eq:ReturnProb}: $\overline{P} = N_{\rm sec}/\text{dim} \mathcal{H}$, $\overline{P}_{\rm max} = 1/D_{\mathcal{H}_{\rm max}} $ and $\overline{P_0} = 1/\text{dim} \mathcal{H}_0$. Since $\text{dim} \mathcal{H}\leq N_{\rm sec}D_{\mathcal{H}_{\rm max}}$, we have $\overline{P}\geq \overline{P_{\rm max}}$, consistent with the findings in Fig.~\ref{fig:ReturnProb}.

%Beyond that, the IPR is useful as a localization probe of a given bitstring as measured in the basis of eigenvectors. For bitstrings that are delocalized over the basis of eigenstates, one can expect the overlaps to behave as $|\langle \boldsymbol{b} | \alpha \rangle|^2 \sim \mathcal{D}_{\text{max}}^{-1}$. As a result, the IPR for such states should scale as $\text{IPR}_{\boldsymbol{b}} \sim \mathcal{D}_{\text{max}}^{-1} $. Vice versa, states localized on a particular set of eigenvectors will show different scaling of IPR with the system size.

\begin{figure}[ ]
\centering
\includegraphics[width=1\columnwidth]{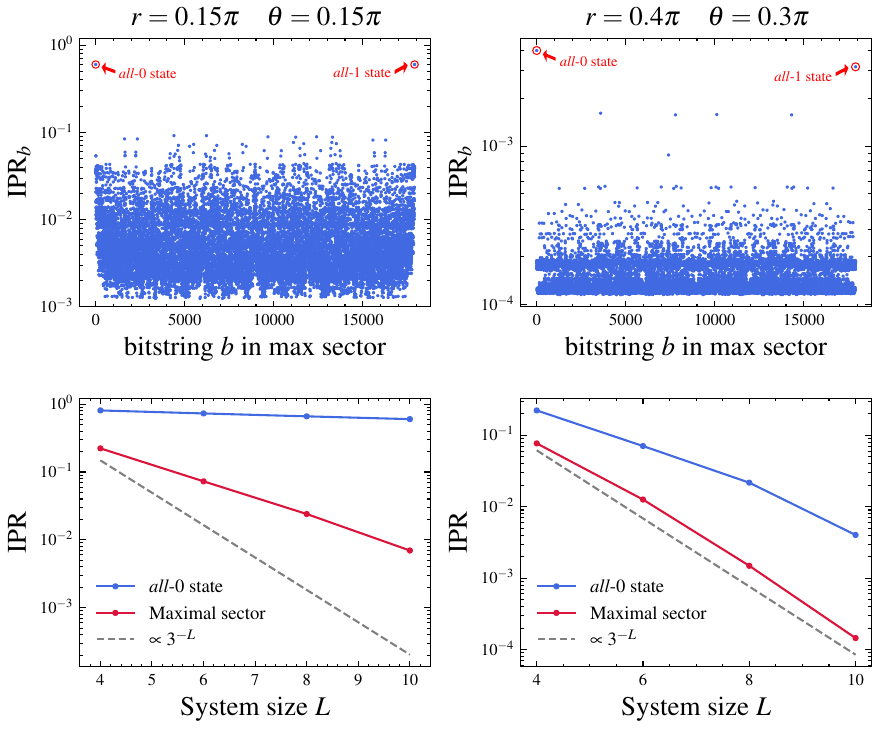}
\caption{ The distribution of the IPR for the basis of bitstring states in the maximal sector for $L=10$ and the scaling of IPR with the system size. The unitary matrix for the PDC gate is $u = e^{ i r (\sin\theta \sigma^{x} + \cos \theta \sigma^z ) }$, but with two different choices of parameters $r$ and $\theta$.}

\label{fig:IPR}
\end{figure}

As Fig.~\ref{fig:IPR} indicates, the \textit{all}-$0$ and \textit{all}-$1$ states indeed show anomalously large values of the IPR compared to all other bitstring states. While they overlap only with a fraction of eigenstates, the total number is still exponential in the system size $L$. This is in contrast to the phenomenon of quantum scars, where persistent revivals of the return probability can be explained by the fact that some bitstring states have large overlaps with a small $O(L)$ tower of scarred eigenstates that are equidistant in energy~\cite{Turner_2018, Serbyn_2021}.

\begin{figure}[ ]
\centering
\includegraphics[width=1.02\columnwidth]{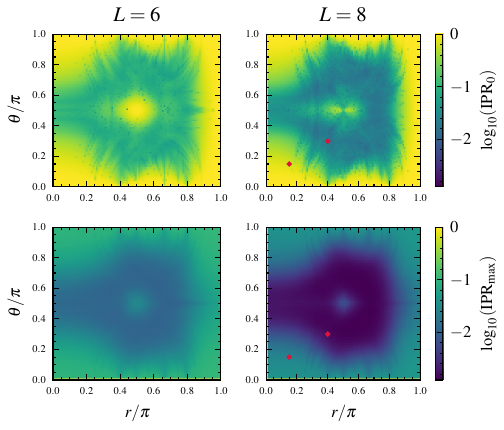}
\caption{The IPR heatmap for the zero state $\text{IPR}_0$ and for the full maximal sector $\text{IPR}_{\text{max}}$ in a range of parameters $r \in[0,\pi]$ and $\theta \in [0,\pi]$ for the unitary $u = e^{ i r (\sin\theta \sigma^{x} + \cos \theta \sigma^z ) }$ of the PDC gate. Red squares are at coordinates $(r,\theta)$ that correspond to parameters from Fig.~\ref{fig:IPR}. }
\label{fig:heatmap}
\end{figure}

Finally, to illustrate that the phenomenon discussed is robust across different parameter regimes, we show in Fig.~\ref{fig:heatmap} heat maps of the IPR for two system sizes, considering both the zero state and the entire maximal sector.

% To complement the IPR analysis, we also produce these heat maps for the spectral statistics via the ratio factor, defined for three consecutive eigenphases ${\varphi_{n-1}, \varphi_n, \varphi_{n+1}}$ as $r_n = \frac{\min(\Delta \varphi_n,\Delta \varphi_{n+1})}{\max(\Delta \varphi_n,\Delta \varphi_{n+1})}$ with $\Delta \varphi_n = \varphi_n - \varphi_{n-1}$ and $0<r_n<1$. Integrable systems have uncorrelated levels described by a Poisson distribution with average ratio factor $\langle r \rangle \approx 0.386$ \cite{atas_distribution_2013}, while chaotic systems are captured by random-matrix theory, with a GUE (or CUE) value of $\langle r \rangle \approx 0.6$ \cite{atas_distribution_2013}. Figure~\ref{fig:ratio_factor} shows heat maps analogous to those studied for the IPR. 

% \begin{figure}[H]
% \centering
% \includegraphics[width=.95\columnwidth]{images_main/Infinite_temp_auto_correlator_Z_r0.63_theta0.94.png}
% \caption{Infinite temperature correlator.}
% \label{fig:Sztotal}
% \end{figure}

% Scaling of the IPR with the system size for \textit{all}-$0$ state and for averaged over all bitstrings in the maximal sector. For both plots the unitary for the PDC gate is $u = e^{ i r (\sin\theta \sigma^{x} + \cos \theta \sigma^z ) }$, but with two different choices of parameters $r$ and $\theta$. Insets: the distribution of IPR for the basis of bitstring states in the maximal sector for a circuit with $L=10$. Red circles correspond to \textit{all}-$0$ and \textit{all}-$1$ states.

% for bitstring states in the maximal sector of ladder PDC circuit. Top Panel: the distribution of IPR for the basis of bitstring states in the maximal sector for a circuit with $L=10$. \textit{All}-$0$ and \textit{all}-$1$ states have anomolosly large value of the IPR. Red dashed line -- mean value of IPR. Bottom panel: scaling of IPR with the system size. While the mean value of IPR seems to behave as $\sim \mathcal{D}_{\text{max}}^{-1}$ , the IPR of the \textit{all}-$0$ state decays much slower. For both plots the unitary for the PDC gate is $u = e^{ i r (\sin\theta \sigma^{x} + \cos \theta \sigma^z ) }$ with $r=0.2 \pi$ and $\theta = 0.3 \pi$.}

% \begin{figure}[ ]
% \centering
% \includegraphics[width=0.95\columnwidth]{images_main/r_mean_L4_L6.png}
% \caption{Heatmap of ratio factor for the PDC gate $u = e^{ i r (\sin\theta \sigma^{x} + \cos \theta \sigma^z ) }$, with random diagonal phases (see \eqref{eq:PDC_gate}).}
% \label{fig:ratio_factor}
% \end{figure}

\subsection{Eigenstate diagnostics II: matrix elements of local density}

To further quantify ergodicity in the maximal sector, we analyzed the diagonal matrix elements of local observables.
% and the statistics of energy phases. 

\begin{figure}[ ]
\centering
\includegraphics[width=0.9\columnwidth]{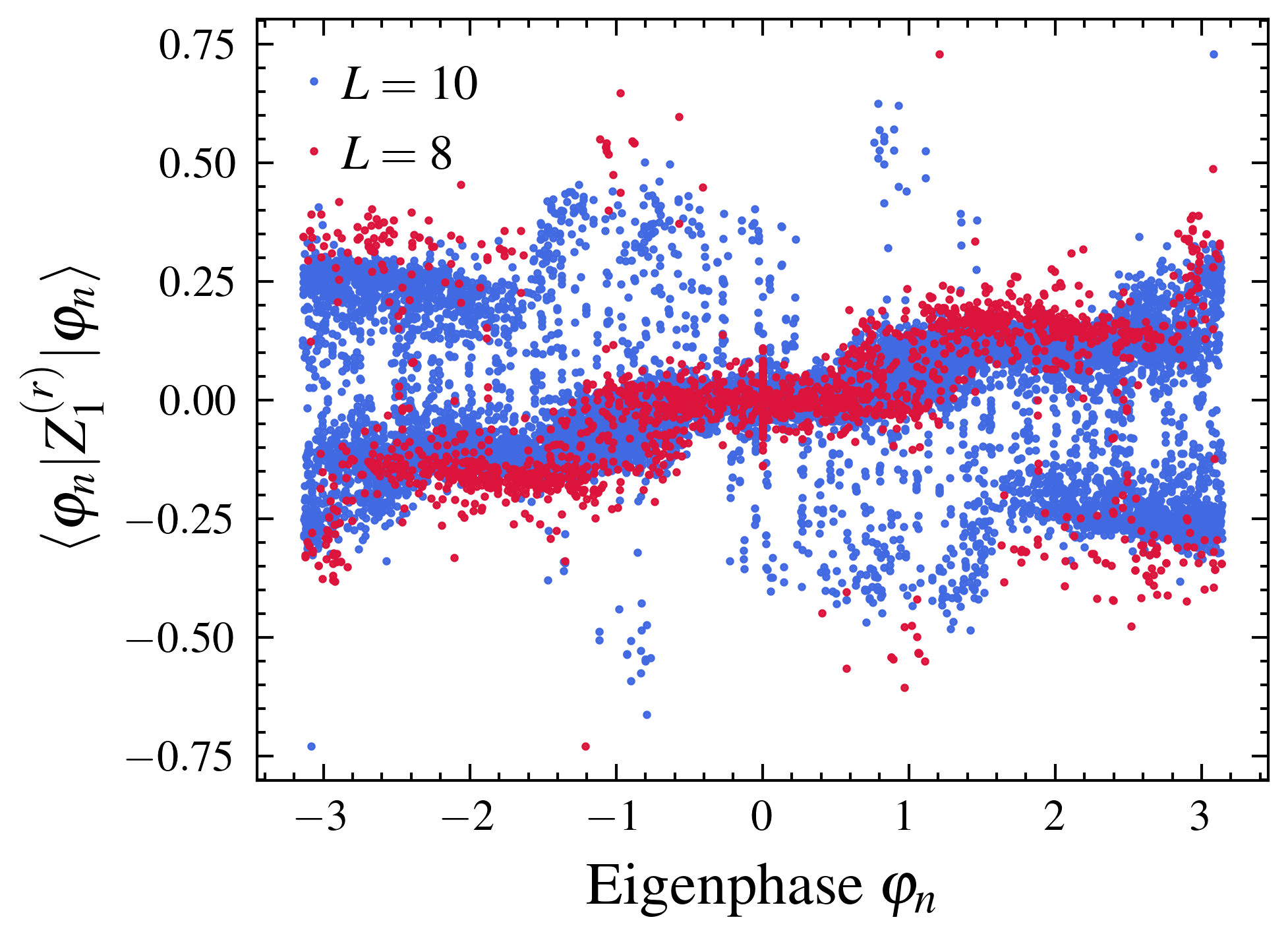}
\caption{Diagonal matrix elements of local observable $Z_{1}^{(r)}$ computed in the eigenbasis of the PDC circuit with the ladder geometry and $L=10$. The unitary for the PDC gate is $u = e^{ i r (\sin\theta \sigma^{x} + \cos \theta \sigma^z ) }$ with $r=0.2 \pi$ and $\theta = 0.3 \pi$. }
\label{fig:matrix-elements}
\end{figure}

In the thermodynamic limit, the diagonal elements of local observables in the eigenbasis of chaotic Hamiltonians vary smoothly with energy density, a behavior known as the diagonal Eigenstate Thermalization Hypothesis (ETH) \cite{Deutsch91, Srednicki94, Deutsch_2018}. In the case of chaotic Floquet dynamics, where energy is not conserved, instead one would expect the diagonal matrix elements to coincide with the prediction of the maximally mixed (infinite temperature) state, $\rho_{\infty} = \frac{1}{\mathcal{D}} \text{Id}$. For example, in the maximal sector, the expectation value of the observable $Z_{1}^{(r)}$ in the infinite-temperature ensemble is zero. However, as shown in Fig.~\ref{fig:matrix-elements}, the diagonal matrix elements do not become smoother nor approach zero as the system size increases. In contrast, a large number of outliers are present, and their number appears to grow with increasing system size.
We note that this violation of ETH is observed for typical parameter values $r,\theta$ far enough from the trivial (integrable) regimes.
\\\\
We close this section with a curious remark. One can replace the unitary gate in (\ref{eq:PDC_gate}) by a stochastic matrix with
$\varphi_j=0$ and $\alpha, \,\beta, \,\gamma=1-\alpha, \, \delta=1-\beta\in\mathbb R^+$, and study the corresponding many-body markov chain PDC circuit. Considering an open-ladder geometry and attaching two stochastic reservoirs at the ends, one notices that the corresponding nonequilibrium steady-state probability vector admits matrix product representation with bond dimension growing only linearly with the distance from the boundary. This is
very similar to stochastic deformation of the rule-54 reversible cellular automaton~\cite{Rule54review,PalettaProsen} and suggests potentially nontrivial integrability structure.

%Consistent with the strong violation of the diagonal ETH, we find that the statistics of eigenphases 
%$\{\varphi_{n}\}_{n=1}^{\mathcal{D}_{\text{max}}}$ do not follow the predictions of Random Matrix Theory \cite{cite}. 
%We numerically computed the spectral form factor (SFF),
%\begin{equation}
%K(t) = \left| \mathrm{Tr}\, \mathcal{U}^{t} \right|^2 = \sum_{m,n} e^{i (\varphi_{m} - \varphi_{n}) t},
%\end{equation}
%with the results shown in Fig.~\ref{fig:sff}. 
%The behavior of the SFF is clearly Poissonian, suggesting that the maximal sector may exhibit integrable dynamics.

%\po{Tomaž wanted to mention the existence of MPS NESS for the maximal sector and boundary drive. Should we or not?}

\section{Discussion and Outlook}

In this work, we introduced a new class of systems, which we termed PDC circuits. These circuits are built from local gates constrained such that, for any pair of degrees of freedom they act on, the gate commutes with the
difference of the local one-body operator, see Eq.(\ref{PDC-condition}). The construction can be extended to degrees of freedom with arbitrary local Hilbert space dimension - and even to purely classical spins, where the condition becomes Poissonian commutativity. Placing the degrees of freedom on the edges of a graph and attaching a gate to each vertex, we obtain circuits endowed with an extensive family of local symmetries: one for every closed even-length walk (loop) on the graph. Although the graphs can be of arbitrary dimension, these symmetries are intrinsically one-dimensional, reminiscent of 1-form symmetries. Nevertheless, while the number of such charges grows extensively with system size, it is not sufficient to render the model integrable (at least in the classical sense).

% We examined in detail several regular geometries, including the square lattice in 2D and the ladder in 1D. In both cases, the number of independent loop symmetries grows extensively with the system size, though not sufficiently to render the model integrable, at least in the classical sense.

As a case study, we analyzed a 1D ladder graph. The loop symmetries induce Hilbert-space fragmentation: the configuration space splits into dynamically disconnected sectors. On the ladder, we show that every sector can be uniquely labeled by their loop charge induced symmetry values, providing a complete set of quantum numbers.
%A key consequence of these symmetries is Hilbert space fragmentation: the configuration space splits into dynamically disconnected sectors. Using the ladder geometry as an example, we showed that every sector can be uniquely labeled by the symmetry values, providing a complete set of quantum numbers. 
Both the number of such sectors and the dimension of the largest sector grow exponentially with the system size, yet more slowly than the full Hilbert space dimension, signaling strong fragmentation.

We also explore the dynamics in the ladder geometry. At infinite temperature, correlation functions develop a nonzero long-time plateau, reflecting fragmentation-induced non-ergodicity. But even within the largest sector — where correlations relax to zero — we still observe signatures of non-ergodicity, manifested through periodic revivals of simple bitstring states and the breakdown of the diagonal ETH. 

% , and Poissonian level statistics.

Our study establishes PDC circuits as a versatile platform for exploring symmetry-induced fragmentation and nonergodic dynamics. Beyond the specific models discussed here, several open directions naturally emerge. An important open question is whether the dynamics within the maximal sector is ultimately related to some form of integrability or whether it hides more subtle forms of non-ergodicity. On the structural side, one may consider generalizations to three-dimensional lattices, or even surface-like conserved charges, thereby broadening the scope of emergent loop symmetries. Connections to paradigmatic models such as the toric code raise the intriguing possibility of exploiting these symmetries for robust quantum memory applications. From a dynamical perspective, it would be interesting to know if we can establish a genuine quantum localization-delocalization transition, % characterize information spreading or localization–delocalization transitions, 
for instance via out-of-time-order correlators (OTOCs), as recently studied in a cellular automata setting of the PDC circuit by means of the classical decorrelator~\cite{Yusuf}. In that vein, investigating transport along topological loops may reveal new out-of-equilibrium behavior. We also note that the explicit lattice construction of 1-form symmetries in this work could potentially inform their construction in continuum QFTs, where direct implementations are considerably harder. Finally, it may be of interest to investigate the stability of these phenomena -- especially the late time revivals of certain bitstring states -- under symmetry-breaking perturbations.

\section*{Acknowledgments}
We are very grateful to Yusuf Kasim for fruitful discussions and collaboration on related projects, and to Zyra Lenny for inspiration. PO gratefully acknowledges Ferme de Champdolent and Didier Saint-Roch for
providing an inspiring environment during the summer, where part of this manuscript was written.
We acknowledge funding from the European Union HORIZON-CL4-2022-QUANTUM-02-SGA through PASQuanS2.1 (Grant Agreement No. 101113690), European Research Council (ERC) through Advanced grant QUEST (Grant Agreement No. 101096208), as well as the Slovenian Research and Innovation agency (ARIS) through the Program P1-0402.

\bibliographystyle{apsrev4-2}
\bibliography{references}

\begin{thebibliography}{51}%
\makeatletter
\providecommand \@ifxundefined [1]{%
 \@ifx{#1\undefined}
}%
\providecommand \@ifnum [1]{%
 \ifnum #1\expandafter \@firstoftwo
 \else \expandafter \@secondoftwo
 \fi
}%
\providecommand \@ifx [1]{%
 \ifx #1\expandafter \@firstoftwo
 \else \expandafter \@secondoftwo
 \fi
}%
\providecommand \natexlab [1]{#1}%
\providecommand \enquote  [1]{``#1''}%
\providecommand \bibnamefont  [1]{#1}%
\providecommand \bibfnamefont [1]{#1}%
\providecommand \citenamefont [1]{#1}%
\providecommand \href@noop [0]{\@secondoftwo}%
\providecommand \href [0]{\begingroup \@sanitize@url \@href}%
\providecommand \@href[1]{\@@startlink{#1}\@@href}%
\providecommand \@@href[1]{\endgroup#1\@@endlink}%
\providecommand \@sanitize@url [0]{\catcode `\\12\catcode `\$12\catcode `\&12\catcode `\#12\catcode `\^12\catcode `\_12\catcode `\%12\relax}%
\providecommand \@@startlink[1]{}%
\providecommand \@@endlink[0]{}%
\providecommand \url  [0]{\begingroup\@sanitize@url \@url }%
\providecommand \@url [1]{\endgroup\@href {#1}{\urlprefix }}%
\providecommand \urlprefix  [0]{URL }%
\providecommand \Eprint [0]{\href }%
\providecommand \doibase [0]{https://doi.org/}%
\providecommand \selectlanguage [0]{\@gobble}%
\providecommand \bibinfo  [0]{\@secondoftwo}%
\providecommand \bibfield  [0]{\@secondoftwo}%
\providecommand \translation [1]{[#1]}%
\providecommand \BibitemOpen [0]{}%
\providecommand \bibitemStop [0]{}%
\providecommand \bibitemNoStop [0]{.\EOS\space}%
\providecommand \EOS [0]{\spacefactor3000\relax}%
\providecommand \BibitemShut  [1]{\csname bibitem#1\endcsname}%
\let\auto@bib@innerbib\@empty
%</preamble>
\bibitem [{\citenamefont {Franchini}(2017)}]{Franchini_2017}%
  \BibitemOpen
  \bibfield  {author} {\bibinfo {author} {\bibfnamefont {F.}~\bibnamefont {Franchini}},\ }\href {https://doi.org/10.1007/978-3-319-48487-7} {\emph {\bibinfo {title} {An Introduction to Integrable Techniques for One-Dimensional Quantum Systems}}}\ (\bibinfo  {publisher} {Springer International Publishing},\ \bibinfo {year} {2017})\BibitemShut {NoStop}%
\bibitem [{\citenamefont {Korepin}\ \emph {et~al.}(1993)\citenamefont {Korepin}, \citenamefont {Izergin},\ and\ \citenamefont {Bogoliubov}}]{korepin1993quantuminversescatteringmethod}%
  \BibitemOpen
  \bibfield  {author} {\bibinfo {author} {\bibfnamefont {V.~E.}\ \bibnamefont {Korepin}}, \bibinfo {author} {\bibfnamefont {A.~G.}\ \bibnamefont {Izergin}},\ and\ \bibinfo {author} {\bibfnamefont {N.~M.}\ \bibnamefont {Bogoliubov}},\ }\href {https://arxiv.org/abs/cond-mat/9301031} {\bibinfo {title} {Quantum inverse scattering method and correlation functions}} (\bibinfo {year} {1993}),\ \Eprint {https://arxiv.org/abs/cond-mat/9301031} {arXiv:cond-mat/9301031 [cond-mat]} \BibitemShut {NoStop}%
\bibitem [{\citenamefont {Caux}\ and\ \citenamefont {Mossel}(2011)}]{Caux_2011}%
  \BibitemOpen
  \bibfield  {author} {\bibinfo {author} {\bibfnamefont {J.-S.}\ \bibnamefont {Caux}}\ and\ \bibinfo {author} {\bibfnamefont {J.}~\bibnamefont {Mossel}},\ }\href {https://doi.org/10.1088/1742-5468/2011/02/p02023} {\bibfield  {journal} {\bibinfo  {journal} {Journal of Statistical Mechanics: Theory and Experiment}\ }\textbf {\bibinfo {volume} {2011}},\ \bibinfo {pages} {P02023} (\bibinfo {year} {2011})}\BibitemShut {NoStop}%
\bibitem [{\citenamefont {Ilievski}\ \emph {et~al.}(2016)\citenamefont {Ilievski}, \citenamefont {Medenjak}, \citenamefont {Prosen},\ and\ \citenamefont {Zadnik}}]{Ilievski_2016}%
  \BibitemOpen
  \bibfield  {author} {\bibinfo {author} {\bibfnamefont {E.}~\bibnamefont {Ilievski}}, \bibinfo {author} {\bibfnamefont {M.}~\bibnamefont {Medenjak}}, \bibinfo {author} {\bibfnamefont {T.}~\bibnamefont {Prosen}},\ and\ \bibinfo {author} {\bibfnamefont {L.}~\bibnamefont {Zadnik}},\ }\href {https://doi.org/10.1088/1742-5468/2016/06/064008} {\bibfield  {journal} {\bibinfo  {journal} {Journal of Statistical Mechanics: Theory and Experiment}\ }\textbf {\bibinfo {volume} {2016}},\ \bibinfo {pages} {064008} (\bibinfo {year} {2016})}\BibitemShut {NoStop}%
\bibitem [{\citenamefont {Caux}\ and\ \citenamefont {Essler}(2013)}]{Caux_2013}%
  \BibitemOpen
  \bibfield  {author} {\bibinfo {author} {\bibfnamefont {J.-S.}\ \bibnamefont {Caux}}\ and\ \bibinfo {author} {\bibfnamefont {F.~H.~L.}\ \bibnamefont {Essler}},\ }\bibfield  {journal} {\bibinfo  {journal} {Physical Review Letters}\ }\textbf {\bibinfo {volume} {110}},\ \href {https://doi.org/10.1103/physrevlett.110.257203} {10.1103/physrevlett.110.257203} (\bibinfo {year} {2013})\BibitemShut {NoStop}%
\bibitem [{\citenamefont {Ilievski}\ \emph {et~al.}(2015)\citenamefont {Ilievski}, \citenamefont {De~Nardis}, \citenamefont {Wouters}, \citenamefont {Caux}, \citenamefont {Essler},\ and\ \citenamefont {Prosen}}]{Ilievski_2015}%
  \BibitemOpen
  \bibfield  {author} {\bibinfo {author} {\bibfnamefont {E.}~\bibnamefont {Ilievski}}, \bibinfo {author} {\bibfnamefont {J.}~\bibnamefont {De~Nardis}}, \bibinfo {author} {\bibfnamefont {B.}~\bibnamefont {Wouters}}, \bibinfo {author} {\bibfnamefont {J.-S.}\ \bibnamefont {Caux}}, \bibinfo {author} {\bibfnamefont {F.}~\bibnamefont {Essler}},\ and\ \bibinfo {author} {\bibfnamefont {T.}~\bibnamefont {Prosen}},\ }\bibfield  {journal} {\bibinfo  {journal} {Physical Review Letters}\ }\textbf {\bibinfo {volume} {115}},\ \href {https://doi.org/10.1103/physrevlett.115.157201} {10.1103/physrevlett.115.157201} (\bibinfo {year} {2015})\BibitemShut {NoStop}%
\bibitem [{\citenamefont {Vidmar}\ and\ \citenamefont {Rigol}(2016)}]{Vidmar_2016}%
  \BibitemOpen
  \bibfield  {author} {\bibinfo {author} {\bibfnamefont {L.}~\bibnamefont {Vidmar}}\ and\ \bibinfo {author} {\bibfnamefont {M.}~\bibnamefont {Rigol}},\ }\href {https://doi.org/10.1088/1742-5468/2016/06/064007} {\bibfield  {journal} {\bibinfo  {journal} {Journal of Statistical Mechanics: Theory and Experiment}\ }\textbf {\bibinfo {volume} {2016}},\ \bibinfo {pages} {064007} (\bibinfo {year} {2016})}\BibitemShut {NoStop}%
\bibitem [{\citenamefont {Essler}\ and\ \citenamefont {Fagotti}(2016)}]{Essler_2016}%
  \BibitemOpen
  \bibfield  {author} {\bibinfo {author} {\bibfnamefont {F.~H.~L.}\ \bibnamefont {Essler}}\ and\ \bibinfo {author} {\bibfnamefont {M.}~\bibnamefont {Fagotti}},\ }\href {https://doi.org/10.1088/1742-5468/2016/06/064002} {\bibfield  {journal} {\bibinfo  {journal} {Journal of Statistical Mechanics: Theory and Experiment}\ }\textbf {\bibinfo {volume} {2016}},\ \bibinfo {pages} {064002} (\bibinfo {year} {2016})}\BibitemShut {NoStop}%
\bibitem [{\citenamefont {Castro-Alvaredo}\ \emph {et~al.}(2016)\citenamefont {Castro-Alvaredo}, \citenamefont {Doyon},\ and\ \citenamefont {Yoshimura}}]{Castro_Alvaredo_2016}%
  \BibitemOpen
  \bibfield  {author} {\bibinfo {author} {\bibfnamefont {O.~A.}\ \bibnamefont {Castro-Alvaredo}}, \bibinfo {author} {\bibfnamefont {B.}~\bibnamefont {Doyon}},\ and\ \bibinfo {author} {\bibfnamefont {T.}~\bibnamefont {Yoshimura}},\ }\bibfield  {journal} {\bibinfo  {journal} {Physical Review X}\ }\textbf {\bibinfo {volume} {6}},\ \href {https://doi.org/10.1103/physrevx.6.041065} {10.1103/physrevx.6.041065} (\bibinfo {year} {2016})\BibitemShut {NoStop}%
\bibitem [{\citenamefont {Bertini}\ \emph {et~al.}(2016)\citenamefont {Bertini}, \citenamefont {Collura}, \citenamefont {De~Nardis},\ and\ \citenamefont {Fagotti}}]{Bertini_2016}%
  \BibitemOpen
  \bibfield  {author} {\bibinfo {author} {\bibfnamefont {B.}~\bibnamefont {Bertini}}, \bibinfo {author} {\bibfnamefont {M.}~\bibnamefont {Collura}}, \bibinfo {author} {\bibfnamefont {J.}~\bibnamefont {De~Nardis}},\ and\ \bibinfo {author} {\bibfnamefont {M.}~\bibnamefont {Fagotti}},\ }\bibfield  {journal} {\bibinfo  {journal} {Physical Review Letters}\ }\textbf {\bibinfo {volume} {117}},\ \href {https://doi.org/10.1103/physrevlett.117.207201} {10.1103/physrevlett.117.207201} (\bibinfo {year} {2016})\BibitemShut {NoStop}%
\bibitem [{\citenamefont {Essler}(2023)}]{Essler_2023}%
  \BibitemOpen
  \bibfield  {author} {\bibinfo {author} {\bibfnamefont {F.~H.}\ \bibnamefont {Essler}},\ }\href {https://doi.org/10.1016/j.physa.2022.127572} {\bibfield  {journal} {\bibinfo  {journal} {Physica A: Statistical Mechanics and its Applications}\ }\textbf {\bibinfo {volume} {631}},\ \bibinfo {pages} {127572} (\bibinfo {year} {2023})}\BibitemShut {NoStop}%
\bibitem [{\citenamefont {Doyon}(2020)}]{Doyon_2020}%
  \BibitemOpen
  \bibfield  {author} {\bibinfo {author} {\bibfnamefont {B.}~\bibnamefont {Doyon}},\ }\bibfield  {journal} {\bibinfo  {journal} {SciPost Physics Lecture Notes}\ }\href {https://doi.org/10.21468/scipostphyslectnotes.18} {10.21468/scipostphyslectnotes.18} (\bibinfo {year} {2020})\BibitemShut {NoStop}%
\bibitem [{\citenamefont {Abanin}\ \emph {et~al.}(2019)\citenamefont {Abanin}, \citenamefont {Altman}, \citenamefont {Bloch},\ and\ \citenamefont {Serbyn}}]{Abanin_2019}%
  \BibitemOpen
  \bibfield  {author} {\bibinfo {author} {\bibfnamefont {D.~A.}\ \bibnamefont {Abanin}}, \bibinfo {author} {\bibfnamefont {E.}~\bibnamefont {Altman}}, \bibinfo {author} {\bibfnamefont {I.}~\bibnamefont {Bloch}},\ and\ \bibinfo {author} {\bibfnamefont {M.}~\bibnamefont {Serbyn}},\ }\bibfield  {journal} {\bibinfo  {journal} {Reviews of Modern Physics}\ }\textbf {\bibinfo {volume} {91}},\ \href {https://doi.org/10.1103/revmodphys.91.021001} {10.1103/revmodphys.91.021001} (\bibinfo {year} {2019})\BibitemShut {NoStop}%
\bibitem [{\citenamefont {Sierant}\ \emph {et~al.}(2025)\citenamefont {Sierant}, \citenamefont {Lewenstein}, \citenamefont {Scardicchio}, \citenamefont {Vidmar},\ and\ \citenamefont {Zakrzewski}}]{Sierant_2025}%
  \BibitemOpen
  \bibfield  {author} {\bibinfo {author} {\bibfnamefont {P.}~\bibnamefont {Sierant}}, \bibinfo {author} {\bibfnamefont {M.}~\bibnamefont {Lewenstein}}, \bibinfo {author} {\bibfnamefont {A.}~\bibnamefont {Scardicchio}}, \bibinfo {author} {\bibfnamefont {L.}~\bibnamefont {Vidmar}},\ and\ \bibinfo {author} {\bibfnamefont {J.}~\bibnamefont {Zakrzewski}},\ }\href {https://doi.org/10.1088/1361-6633/ad9756} {\bibfield  {journal} {\bibinfo  {journal} {Reports on Progress in Physics}\ }\textbf {\bibinfo {volume} {88}},\ \bibinfo {pages} {026502} (\bibinfo {year} {2025})}\BibitemShut {NoStop}%
\bibitem [{\citenamefont {Serbyn}\ \emph {et~al.}(2013)\citenamefont {Serbyn}, \citenamefont {Papić},\ and\ \citenamefont {Abanin}}]{Serbyn_2013}%
  \BibitemOpen
  \bibfield  {author} {\bibinfo {author} {\bibfnamefont {M.}~\bibnamefont {Serbyn}}, \bibinfo {author} {\bibfnamefont {Z.}~\bibnamefont {Papić}},\ and\ \bibinfo {author} {\bibfnamefont {D.~A.}\ \bibnamefont {Abanin}},\ }\bibfield  {journal} {\bibinfo  {journal} {Physical Review Letters}\ }\textbf {\bibinfo {volume} {111}},\ \href {https://doi.org/10.1103/physrevlett.111.127201} {10.1103/physrevlett.111.127201} (\bibinfo {year} {2013})\BibitemShut {NoStop}%
\bibitem [{\citenamefont {Imbrie}\ \emph {et~al.}(2017)\citenamefont {Imbrie}, \citenamefont {Ros},\ and\ \citenamefont {Scardicchio}}]{Imbrie_2017}%
  \BibitemOpen
  \bibfield  {author} {\bibinfo {author} {\bibfnamefont {J.~Z.}\ \bibnamefont {Imbrie}}, \bibinfo {author} {\bibfnamefont {V.}~\bibnamefont {Ros}},\ and\ \bibinfo {author} {\bibfnamefont {A.}~\bibnamefont {Scardicchio}},\ }\bibfield  {journal} {\bibinfo  {journal} {Annalen der Physik}\ }\textbf {\bibinfo {volume} {529}},\ \href {https://doi.org/10.1002/andp.201600278} {10.1002/andp.201600278} (\bibinfo {year} {2017})\BibitemShut {NoStop}%
\bibitem [{\citenamefont {Khemani}\ \emph {et~al.}(2020)\citenamefont {Khemani}, \citenamefont {Hermele},\ and\ \citenamefont {Nandkishore}}]{Khemani_2020}%
  \BibitemOpen
  \bibfield  {author} {\bibinfo {author} {\bibfnamefont {V.}~\bibnamefont {Khemani}}, \bibinfo {author} {\bibfnamefont {M.}~\bibnamefont {Hermele}},\ and\ \bibinfo {author} {\bibfnamefont {R.}~\bibnamefont {Nandkishore}},\ }\bibfield  {journal} {\bibinfo  {journal} {Physical Review B}\ }\textbf {\bibinfo {volume} {101}},\ \href {https://doi.org/10.1103/physrevb.101.174204} {10.1103/physrevb.101.174204} (\bibinfo {year} {2020})\BibitemShut {NoStop}%
\bibitem [{\citenamefont {Sala}\ \emph {et~al.}(2020{\natexlab{a}})\citenamefont {Sala}, \citenamefont {Rakovszky}, \citenamefont {Verresen}, \citenamefont {Knap},\ and\ \citenamefont {Pollmann}}]{DCmodels}%
  \BibitemOpen
  \bibfield  {author} {\bibinfo {author} {\bibfnamefont {P.}~\bibnamefont {Sala}}, \bibinfo {author} {\bibfnamefont {T.}~\bibnamefont {Rakovszky}}, \bibinfo {author} {\bibfnamefont {R.}~\bibnamefont {Verresen}}, \bibinfo {author} {\bibfnamefont {M.}~\bibnamefont {Knap}},\ and\ \bibinfo {author} {\bibfnamefont {F.}~\bibnamefont {Pollmann}},\ }\href {https://doi.org/10.1103/PhysRevX.10.011047} {\bibfield  {journal} {\bibinfo  {journal} {Phys. Rev. X}\ }\textbf {\bibinfo {volume} {10}},\ \bibinfo {pages} {011047} (\bibinfo {year} {2020}{\natexlab{a}})}\BibitemShut {NoStop}%
\bibitem [{\citenamefont {Zadnik}\ and\ \citenamefont {Fagotti}(2021)}]{zadnik2021folded}%
  \BibitemOpen
  \bibfield  {author} {\bibinfo {author} {\bibfnamefont {L.}~\bibnamefont {Zadnik}}\ and\ \bibinfo {author} {\bibfnamefont {M.}~\bibnamefont {Fagotti}},\ }\href@noop {} {\bibfield  {journal} {\bibinfo  {journal} {SciPost Physics Core}\ }\textbf {\bibinfo {volume} {4}},\ \bibinfo {pages} {010} (\bibinfo {year} {2021})}\BibitemShut {NoStop}%
\bibitem [{\citenamefont {Moudgalya}\ \emph {et~al.}(2022)\citenamefont {Moudgalya}, \citenamefont {Bernevig},\ and\ \citenamefont {Regnault}}]{Moudgalya_2022}%
  \BibitemOpen
  \bibfield  {author} {\bibinfo {author} {\bibfnamefont {S.}~\bibnamefont {Moudgalya}}, \bibinfo {author} {\bibfnamefont {B.~A.}\ \bibnamefont {Bernevig}},\ and\ \bibinfo {author} {\bibfnamefont {N.}~\bibnamefont {Regnault}},\ }\href {https://doi.org/10.1088/1361-6633/ac73a0} {\bibfield  {journal} {\bibinfo  {journal} {Reports on Progress in Physics}\ }\textbf {\bibinfo {volume} {85}},\ \bibinfo {pages} {086501} (\bibinfo {year} {2022})}\BibitemShut {NoStop}%
\bibitem [{\citenamefont {Zadnik}\ and\ \citenamefont {Garrahan}(2023)}]{zadnik2023slow}%
  \BibitemOpen
  \bibfield  {author} {\bibinfo {author} {\bibfnamefont {L.}~\bibnamefont {Zadnik}}\ and\ \bibinfo {author} {\bibfnamefont {J.~P.}\ \bibnamefont {Garrahan}},\ }\href {https://arxiv.org/abs/2304.10394} {\bibinfo {title} {Slow heterogeneous relaxation due to constraints in dual xxz models}} (\bibinfo {year} {2023}),\ \Eprint {https://arxiv.org/abs/2304.10394} {arXiv:2304.10394 [cond-mat.stat-mech]} \BibitemShut {NoStop}%
\bibitem [{\citenamefont {Marić}\ \emph {et~al.}(2025)\citenamefont {Marić}, \citenamefont {Paljk},\ and\ \citenamefont {Zadnik}}]{marić2025slow}%
  \BibitemOpen
  \bibfield  {author} {\bibinfo {author} {\bibfnamefont {V.}~\bibnamefont {Marić}}, \bibinfo {author} {\bibfnamefont {L.}~\bibnamefont {Paljk}},\ and\ \bibinfo {author} {\bibfnamefont {L.}~\bibnamefont {Zadnik}},\ }\href {https://arxiv.org/abs/2510.03159} {\bibinfo {title} {Slow dynamics from a nested hierarchy of frozen states}} (\bibinfo {year} {2025}),\ \Eprint {https://arxiv.org/abs/2510.03159} {arXiv:2510.03159 [cond-mat.stat-mech]} \BibitemShut {NoStop}%
\bibitem [{\citenamefont {De~Roeck}\ and\ \citenamefont {Huveneers}(2017)}]{De_Roeck_2017}%
  \BibitemOpen
  \bibfield  {author} {\bibinfo {author} {\bibfnamefont {W.}~\bibnamefont {De~Roeck}}\ and\ \bibinfo {author} {\bibfnamefont {F.}~\bibnamefont {Huveneers}},\ }\bibfield  {journal} {\bibinfo  {journal} {Physical Review B}\ }\textbf {\bibinfo {volume} {95}},\ \href {https://doi.org/10.1103/physrevb.95.155129} {10.1103/physrevb.95.155129} (\bibinfo {year} {2017})\BibitemShut {NoStop}%
\bibitem [{\citenamefont {Khudorozhkov}\ \emph {et~al.}(2022)\citenamefont {Khudorozhkov}, \citenamefont {Tiwari}, \citenamefont {Chamon},\ and\ \citenamefont {Neupert}}]{Khudorozhkov_2022}%
  \BibitemOpen
  \bibfield  {author} {\bibinfo {author} {\bibfnamefont {A.}~\bibnamefont {Khudorozhkov}}, \bibinfo {author} {\bibfnamefont {A.}~\bibnamefont {Tiwari}}, \bibinfo {author} {\bibfnamefont {C.}~\bibnamefont {Chamon}},\ and\ \bibinfo {author} {\bibfnamefont {T.}~\bibnamefont {Neupert}},\ }\bibfield  {journal} {\bibinfo  {journal} {SciPost Physics}\ }\textbf {\bibinfo {volume} {13}},\ \href {https://doi.org/10.21468/scipostphys.13.4.098} {10.21468/scipostphys.13.4.098} (\bibinfo {year} {2022})\BibitemShut {NoStop}%
\bibitem [{\citenamefont {Ciavarella}\ \emph {et~al.}(2025)\citenamefont {Ciavarella}, \citenamefont {Bauer},\ and\ \citenamefont {Halimeh}}]{ciavarella2025}%
  \BibitemOpen
  \bibfield  {author} {\bibinfo {author} {\bibfnamefont {A.~N.}\ \bibnamefont {Ciavarella}}, \bibinfo {author} {\bibfnamefont {C.~W.}\ \bibnamefont {Bauer}},\ and\ \bibinfo {author} {\bibfnamefont {J.~C.}\ \bibnamefont {Halimeh}},\ }\href {https://arxiv.org/abs/2502.03533} {\bibinfo {title} {Generic hilbert space fragmentation in kogut--susskind lattice gauge theories}} (\bibinfo {year} {2025}),\ \Eprint {https://arxiv.org/abs/2502.03533} {arXiv:2502.03533 [quant-ph]} \BibitemShut {NoStop}%
\bibitem [{\citenamefont {Adler}\ \emph {et~al.}(2024)\citenamefont {Adler}, \citenamefont {Wei}, \citenamefont {Will}, \citenamefont {Srakaew}, \citenamefont {Agrawal}, \citenamefont {Weckesser}, \citenamefont {Moessner}, \citenamefont {Pollmann}, \citenamefont {Bloch},\ and\ \citenamefont {Zeiher}}]{adler2024}%
  \BibitemOpen
  \bibfield  {author} {\bibinfo {author} {\bibfnamefont {D.}~\bibnamefont {Adler}}, \bibinfo {author} {\bibfnamefont {D.}~\bibnamefont {Wei}}, \bibinfo {author} {\bibfnamefont {M.}~\bibnamefont {Will}}, \bibinfo {author} {\bibfnamefont {K.}~\bibnamefont {Srakaew}}, \bibinfo {author} {\bibfnamefont {S.}~\bibnamefont {Agrawal}}, \bibinfo {author} {\bibfnamefont {P.}~\bibnamefont {Weckesser}}, \bibinfo {author} {\bibfnamefont {R.}~\bibnamefont {Moessner}}, \bibinfo {author} {\bibfnamefont {F.}~\bibnamefont {Pollmann}}, \bibinfo {author} {\bibfnamefont {I.}~\bibnamefont {Bloch}},\ and\ \bibinfo {author} {\bibfnamefont {J.}~\bibnamefont {Zeiher}},\ }\href {https://arxiv.org/abs/2404.14896} {\bibinfo {title} {Observation of hilbert-space fragmentation and fractonic excitations in two-dimensional hubbard systems}} (\bibinfo {year} {2024}),\ \Eprint {https://arxiv.org/abs/2404.14896} {arXiv:2404.14896 [cond-mat.quant-gas]} \BibitemShut {NoStop}%
\bibitem [{\citenamefont {Kwan}\ \emph {et~al.}(2023)\citenamefont {Kwan}, \citenamefont {Wilhelm}, \citenamefont {Biswas},\ and\ \citenamefont {Parameswaran}}]{kwan2023}%
  \BibitemOpen
  \bibfield  {author} {\bibinfo {author} {\bibfnamefont {Y.~H.}\ \bibnamefont {Kwan}}, \bibinfo {author} {\bibfnamefont {P.~H.}\ \bibnamefont {Wilhelm}}, \bibinfo {author} {\bibfnamefont {S.}~\bibnamefont {Biswas}},\ and\ \bibinfo {author} {\bibfnamefont {S.~A.}\ \bibnamefont {Parameswaran}},\ }\href {https://arxiv.org/abs/2304.02669} {\bibinfo {title} {Minimal hubbard models of maximal hilbert space fragmentation}} (\bibinfo {year} {2023}),\ \Eprint {https://arxiv.org/abs/2304.02669} {arXiv:2304.02669 [cond-mat.stat-mech]} \BibitemShut {NoStop}%
\bibitem [{\citenamefont {Stahl}\ \emph {et~al.}(2025)\citenamefont {Stahl}, \citenamefont {Hart}, \citenamefont {Khudorozhkov},\ and\ \citenamefont {Nandkishore}}]{stahl2025}%
  \BibitemOpen
  \bibfield  {author} {\bibinfo {author} {\bibfnamefont {C.}~\bibnamefont {Stahl}}, \bibinfo {author} {\bibfnamefont {O.}~\bibnamefont {Hart}}, \bibinfo {author} {\bibfnamefont {A.}~\bibnamefont {Khudorozhkov}},\ and\ \bibinfo {author} {\bibfnamefont {R.}~\bibnamefont {Nandkishore}},\ }\href {https://arxiv.org/abs/2505.15889} {\bibinfo {title} {Strong hilbert space fragmentation and fractons from subsystem and higher-form symmetries}} (\bibinfo {year} {2025}),\ \Eprint {https://arxiv.org/abs/2505.15889} {arXiv:2505.15889 [cond-mat.stat-mech]} \BibitemShut {NoStop}%
\bibitem [{\citenamefont {Gaiotto}\ \emph {et~al.}(2015)\citenamefont {Gaiotto}, \citenamefont {Kapustin}, \citenamefont {Seiberg},\ and\ \citenamefont {Willett}}]{Gaiotto_2015}%
  \BibitemOpen
  \bibfield  {author} {\bibinfo {author} {\bibfnamefont {D.}~\bibnamefont {Gaiotto}}, \bibinfo {author} {\bibfnamefont {A.}~\bibnamefont {Kapustin}}, \bibinfo {author} {\bibfnamefont {N.}~\bibnamefont {Seiberg}},\ and\ \bibinfo {author} {\bibfnamefont {B.}~\bibnamefont {Willett}},\ }\bibfield  {journal} {\bibinfo  {journal} {Journal of High Energy Physics}\ }\textbf {\bibinfo {volume} {2015}},\ \href {https://doi.org/10.1007/jhep02(2015)172} {10.1007/jhep02(2015)172} (\bibinfo {year} {2015})\BibitemShut {NoStop}%
\bibitem [{\citenamefont {McGreevy}(2023)}]{McGreevy_2023}%
  \BibitemOpen
  \bibfield  {author} {\bibinfo {author} {\bibfnamefont {J.}~\bibnamefont {McGreevy}},\ }\href {https://doi.org/10.1146/annurev-conmatphys-040721-021029} {\bibfield  {journal} {\bibinfo  {journal} {Annual Review of Condensed Matter Physics}\ }\textbf {\bibinfo {volume} {14}},\ \bibinfo {pages} {57–82} (\bibinfo {year} {2023})}\BibitemShut {NoStop}%
\bibitem [{Note1()}]{Note1}%
  \BibitemOpen
  \bibinfo {note} {Note, however, that anything discussed below can be generalized straightforwardly to qudits of arbitrary local dimension}\BibitemShut {NoStop}%
\bibitem [{\citenamefont {Kasim}\ and\ \citenamefont {Prosen}(2024)}]{kasim2024deterministicmanybodydynamicsmultifractal}%
  \BibitemOpen
  \bibfield  {author} {\bibinfo {author} {\bibfnamefont {Y.}~\bibnamefont {Kasim}}\ and\ \bibinfo {author} {\bibfnamefont {T.}~\bibnamefont {Prosen}},\ }\href {https://arxiv.org/abs/2411.19779} {\bibfield  {journal} {\bibinfo  {journal} {arXiv preprint arXiv:2411.19779}\ } (\bibinfo {year} {2024})}\BibitemShut {NoStop}%
\bibitem [{Note2()}]{Note2}%
  \BibitemOpen
  \bibinfo {note} {More generally, for any closed walk, since vertices and edges may repeat.}\BibitemShut {Stop}%
\bibitem [{Note3()}]{Note3}%
  \BibitemOpen
  \bibinfo {note} {Even though we assume here periodic boundary conditions, the choice of open boundaries does not affect the structure of the loop charges.}\BibitemShut {Stop}%
\bibitem [{\citenamefont {Kitaev}(2003)}]{Kitaev_2003}%
  \BibitemOpen
  \bibfield  {author} {\bibinfo {author} {\bibfnamefont {A.}~\bibnamefont {Kitaev}},\ }\href {https://doi.org/10.1016/s0003-4916(02)00018-0} {\bibfield  {journal} {\bibinfo  {journal} {Annals of Physics}\ }\textbf {\bibinfo {volume} {303}},\ \bibinfo {pages} {2–30} (\bibinfo {year} {2003})}\BibitemShut {NoStop}%
\bibitem [{\citenamefont {Nayak}\ \emph {et~al.}(2008)\citenamefont {Nayak}, \citenamefont {Simon}, \citenamefont {Stern}, \citenamefont {Freedman},\ and\ \citenamefont {Das~Sarma}}]{Nayak_2008}%
  \BibitemOpen
  \bibfield  {author} {\bibinfo {author} {\bibfnamefont {C.}~\bibnamefont {Nayak}}, \bibinfo {author} {\bibfnamefont {S.~H.}\ \bibnamefont {Simon}}, \bibinfo {author} {\bibfnamefont {A.}~\bibnamefont {Stern}}, \bibinfo {author} {\bibfnamefont {M.}~\bibnamefont {Freedman}},\ and\ \bibinfo {author} {\bibfnamefont {S.}~\bibnamefont {Das~Sarma}},\ }\href {https://doi.org/10.1103/revmodphys.80.1083} {\bibfield  {journal} {\bibinfo  {journal} {Reviews of Modern Physics}\ }\textbf {\bibinfo {volume} {80}},\ \bibinfo {pages} {1083–1159} (\bibinfo {year} {2008})}\BibitemShut {NoStop}%
\bibitem [{\citenamefont {Kasim}\ \emph {et~al.}(2025)\citenamefont {Kasim}, \citenamefont {Orlov},\ and\ \citenamefont {Prosen}}]{Yusuf}%
  \BibitemOpen
  \bibfield  {author} {\bibinfo {author} {\bibfnamefont {Y.}~\bibnamefont {Kasim}}, \bibinfo {author} {\bibfnamefont {P.}~\bibnamefont {Orlov}},\ and\ \bibinfo {author} {\bibfnamefont {T.}~\bibnamefont {Prosen}},\ }\href {https://arxiv.org/abs/2509.22368} {\bibinfo {title} {Phase transition from localization to chaos in classical many-body system}} (\bibinfo {year} {2025}),\ \Eprint {https://arxiv.org/abs/2509.22368} {arXiv:2509.22368 [cond-mat.stat-mech]} \BibitemShut {NoStop}%
\bibitem [{\citenamefont {Žnidarič}\ \emph {et~al.}(2025)\citenamefont {Žnidarič}, \citenamefont {Duh},\ and\ \citenamefont {Zadnik}}]{Znidaric_2025}%
  \BibitemOpen
  \bibfield  {author} {\bibinfo {author} {\bibfnamefont {M.}~\bibnamefont {Žnidarič}}, \bibinfo {author} {\bibfnamefont {U.}~\bibnamefont {Duh}},\ and\ \bibinfo {author} {\bibfnamefont {L.}~\bibnamefont {Zadnik}},\ }\bibfield  {journal} {\bibinfo  {journal} {Physical Review B}\ }\textbf {\bibinfo {volume} {112}},\ \href {https://doi.org/10.1103/tqy8-ynpd} {10.1103/tqy8-ynpd} (\bibinfo {year} {2025})\BibitemShut {NoStop}%
\bibitem [{\citenamefont {Pai}\ \emph {et~al.}(2019)\citenamefont {Pai}, \citenamefont {Pretko},\ and\ \citenamefont {Nandkishore}}]{Pai_2019}%
  \BibitemOpen
  \bibfield  {author} {\bibinfo {author} {\bibfnamefont {S.}~\bibnamefont {Pai}}, \bibinfo {author} {\bibfnamefont {M.}~\bibnamefont {Pretko}},\ and\ \bibinfo {author} {\bibfnamefont {R.~M.}\ \bibnamefont {Nandkishore}},\ }\bibfield  {journal} {\bibinfo  {journal} {Physical Review X}\ }\textbf {\bibinfo {volume} {9}},\ \href {https://doi.org/10.1103/physrevx.9.021003} {10.1103/physrevx.9.021003} (\bibinfo {year} {2019})\BibitemShut {NoStop}%
\bibitem [{\citenamefont {Sala}\ \emph {et~al.}(2020{\natexlab{b}})\citenamefont {Sala}, \citenamefont {Rakovszky}, \citenamefont {Verresen}, \citenamefont {Knap},\ and\ \citenamefont {Pollmann}}]{Sala_2020}%
  \BibitemOpen
  \bibfield  {author} {\bibinfo {author} {\bibfnamefont {P.}~\bibnamefont {Sala}}, \bibinfo {author} {\bibfnamefont {T.}~\bibnamefont {Rakovszky}}, \bibinfo {author} {\bibfnamefont {R.}~\bibnamefont {Verresen}}, \bibinfo {author} {\bibfnamefont {M.}~\bibnamefont {Knap}},\ and\ \bibinfo {author} {\bibfnamefont {F.}~\bibnamefont {Pollmann}},\ }\bibfield  {journal} {\bibinfo  {journal} {Physical Review X}\ }\textbf {\bibinfo {volume} {10}},\ \href {https://doi.org/10.1103/physrevx.10.011047} {10.1103/physrevx.10.011047} (\bibinfo {year} {2020}{\natexlab{b}})\BibitemShut {NoStop}%
\bibitem [{\citenamefont {Rakovszky}\ \emph {et~al.}(2020)\citenamefont {Rakovszky}, \citenamefont {Sala}, \citenamefont {Verresen}, \citenamefont {Knap},\ and\ \citenamefont {Pollmann}}]{Rakovszky_2020}%
  \BibitemOpen
  \bibfield  {author} {\bibinfo {author} {\bibfnamefont {T.}~\bibnamefont {Rakovszky}}, \bibinfo {author} {\bibfnamefont {P.}~\bibnamefont {Sala}}, \bibinfo {author} {\bibfnamefont {R.}~\bibnamefont {Verresen}}, \bibinfo {author} {\bibfnamefont {M.}~\bibnamefont {Knap}},\ and\ \bibinfo {author} {\bibfnamefont {F.}~\bibnamefont {Pollmann}},\ }\bibfield  {journal} {\bibinfo  {journal} {Physical Review B}\ }\textbf {\bibinfo {volume} {101}},\ \href {https://doi.org/10.1103/physrevb.101.125126} {10.1103/physrevb.101.125126} (\bibinfo {year} {2020})\BibitemShut {NoStop}%
\bibitem [{\citenamefont {Santos}\ and\ \citenamefont {Torres-Herrera}(2017)}]{santos}%
  \BibitemOpen
  \bibfield  {author} {\bibinfo {author} {\bibfnamefont {L.~F.}\ \bibnamefont {Santos}}\ and\ \bibinfo {author} {\bibfnamefont {E.~J.}\ \bibnamefont {Torres-Herrera}},\ }\href {https://doi.org/10.1063/1.5016140} {\bibfield  {journal} {\bibinfo  {journal} {AIP Conference Proceedings}\ }\textbf {\bibinfo {volume} {1912}},\ \bibinfo {pages} {020015} (\bibinfo {year} {2017})}\BibitemShut {NoStop}%
\bibitem [{\citenamefont {Torres-Herrera}\ \emph {et~al.}(2018)\citenamefont {Torres-Herrera}, \citenamefont {García-García},\ and\ \citenamefont {Santos}}]{Torres_Herrera_2018}%
  \BibitemOpen
  \bibfield  {author} {\bibinfo {author} {\bibfnamefont {E.~J.}\ \bibnamefont {Torres-Herrera}}, \bibinfo {author} {\bibfnamefont {A.~M.}\ \bibnamefont {García-García}},\ and\ \bibinfo {author} {\bibfnamefont {L.~F.}\ \bibnamefont {Santos}},\ }\bibfield  {journal} {\bibinfo  {journal} {Physical Review B}\ }\textbf {\bibinfo {volume} {97}},\ \href {https://doi.org/10.1103/physrevb.97.060303} {10.1103/physrevb.97.060303} (\bibinfo {year} {2018})\BibitemShut {NoStop}%
\bibitem [{\citenamefont {Hopjan}\ and\ \citenamefont {Vidmar}(2023)}]{Hopjan_2023}%
  \BibitemOpen
  \bibfield  {author} {\bibinfo {author} {\bibfnamefont {M.}~\bibnamefont {Hopjan}}\ and\ \bibinfo {author} {\bibfnamefont {L.}~\bibnamefont {Vidmar}},\ }\bibfield  {journal} {\bibinfo  {journal} {Physical Review Letters}\ }\textbf {\bibinfo {volume} {131}},\ \href {https://doi.org/10.1103/physrevlett.131.060404} {10.1103/physrevlett.131.060404} (\bibinfo {year} {2023})\BibitemShut {NoStop}%
\bibitem [{\citenamefont {Turner}\ \emph {et~al.}(2018)\citenamefont {Turner}, \citenamefont {Michailidis}, \citenamefont {Abanin}, \citenamefont {Serbyn},\ and\ \citenamefont {Papić}}]{Turner_2018}%
  \BibitemOpen
  \bibfield  {author} {\bibinfo {author} {\bibfnamefont {C.~J.}\ \bibnamefont {Turner}}, \bibinfo {author} {\bibfnamefont {A.~A.}\ \bibnamefont {Michailidis}}, \bibinfo {author} {\bibfnamefont {D.~A.}\ \bibnamefont {Abanin}}, \bibinfo {author} {\bibfnamefont {M.}~\bibnamefont {Serbyn}},\ and\ \bibinfo {author} {\bibfnamefont {Z.}~\bibnamefont {Papić}},\ }\href {https://doi.org/10.1038/s41567-018-0137-5} {\bibfield  {journal} {\bibinfo  {journal} {Nature Physics}\ }\textbf {\bibinfo {volume} {14}},\ \bibinfo {pages} {745–749} (\bibinfo {year} {2018})}\BibitemShut {NoStop}%
\bibitem [{\citenamefont {Serbyn}\ \emph {et~al.}(2021)\citenamefont {Serbyn}, \citenamefont {Abanin},\ and\ \citenamefont {Papić}}]{Serbyn_2021}%
  \BibitemOpen
  \bibfield  {author} {\bibinfo {author} {\bibfnamefont {M.}~\bibnamefont {Serbyn}}, \bibinfo {author} {\bibfnamefont {D.~A.}\ \bibnamefont {Abanin}},\ and\ \bibinfo {author} {\bibfnamefont {Z.}~\bibnamefont {Papić}},\ }\href {https://doi.org/10.1038/s41567-021-01230-2} {\bibfield  {journal} {\bibinfo  {journal} {Nature Physics}\ }\textbf {\bibinfo {volume} {17}},\ \bibinfo {pages} {675–685} (\bibinfo {year} {2021})}\BibitemShut {NoStop}%
\bibitem [{\citenamefont {Deutsch}(1991)}]{Deutsch91}%
  \BibitemOpen
  \bibfield  {author} {\bibinfo {author} {\bibfnamefont {J.~M.}\ \bibnamefont {Deutsch}},\ }\href {https://doi.org/10.1103/PhysRevA.43.2046} {\bibfield  {journal} {\bibinfo  {journal} {Phys. Rev. A}\ }\textbf {\bibinfo {volume} {43}},\ \bibinfo {pages} {2046} (\bibinfo {year} {1991})}\BibitemShut {NoStop}%
\bibitem [{\citenamefont {Srednicki}(1994)}]{Srednicki94}%
  \BibitemOpen
  \bibfield  {author} {\bibinfo {author} {\bibfnamefont {M.}~\bibnamefont {Srednicki}},\ }\href {https://doi.org/10.1103/PhysRevE.50.888} {\bibfield  {journal} {\bibinfo  {journal} {Phys. Rev. E}\ }\textbf {\bibinfo {volume} {50}},\ \bibinfo {pages} {888} (\bibinfo {year} {1994})}\BibitemShut {NoStop}%
\bibitem [{\citenamefont {Deutsch}(2018)}]{Deutsch_2018}%
  \BibitemOpen
  \bibfield  {author} {\bibinfo {author} {\bibfnamefont {J.~M.}\ \bibnamefont {Deutsch}},\ }\href {https://doi.org/10.1088/1361-6633/aac9f1} {\bibfield  {journal} {\bibinfo  {journal} {Reports on Progress in Physics}\ }\textbf {\bibinfo {volume} {81}},\ \bibinfo {pages} {082001} (\bibinfo {year} {2018})}\BibitemShut {NoStop}%
\bibitem [{\citenamefont {Buča}\ \emph {et~al.}(2021)\citenamefont {Buča}, \citenamefont {Klobas},\ and\ \citenamefont {Prosen}}]{Rule54review}%
  \BibitemOpen
  \bibfield  {author} {\bibinfo {author} {\bibfnamefont {B.}~\bibnamefont {Buča}}, \bibinfo {author} {\bibfnamefont {K.}~\bibnamefont {Klobas}},\ and\ \bibinfo {author} {\bibfnamefont {T.}~\bibnamefont {Prosen}},\ }\href {https://doi.org/10.1088/1742-5468/ac096b} {\bibfield  {journal} {\bibinfo  {journal} {Journal of Statistical Mechanics: Theory and Experiment}\ }\textbf {\bibinfo {volume} {2021}},\ \bibinfo {pages} {074001} (\bibinfo {year} {2021})}\BibitemShut {NoStop}%
\bibitem [{\citenamefont {Paletta}\ and\ \citenamefont {Prosen}(2025)}]{PalettaProsen}%
  \BibitemOpen
  \bibfield  {author} {\bibinfo {author} {\bibfnamefont {C.}~\bibnamefont {Paletta}}\ and\ \bibinfo {author} {\bibfnamefont {T.}~\bibnamefont {Prosen}},\ }\href@noop {} {\bibinfo {title} {On the integrability structure of the deformed rule 54 reversible cellular automaton}} (\bibinfo {year} {2025}),\ \Eprint {https://arxiv.org/abs/2510.XXXXX} {arXiv:2510.XXXXX [cond-mat.stat-mech]} \BibitemShut {NoStop}%
\end{thebibliography}%

\end{document}